\documentclass[floatfix,superscriptaddress,showpacs,amssymb,10pt,aps,prd,reprint,longbibliography]{revtex4-1}

\usepackage{graphicx,epsfig,amssymb} 
\usepackage{amsmath,amsfonts, times}
\usepackage{bm} 

\usepackage[linktocpage,colorlinks]{hyperref}
\usepackage[caption=false]{subfig}
\usepackage[usenames]{color} 
\usepackage{natbib}
\usepackage{soul}

\usepackage[utf8]{inputenc}
\usepackage{float}

\usepackage{booktabs}

\definecolor{coolblack}{rgb}{0.0, 0.18, 0.39}
\definecolor{darkred}{rgb}{0.5,0,0}
\definecolor{darkgreen}{rgb}{0,0.5,0}
\definecolor{darkblue}{rgb}{0,0,0.5}
\definecolor{lapislazuli}{rgb}{0.15, 0.38, 0.61}
\definecolor{venetianred}{rgb}{0.78, 0.03, 0.08}
\definecolor{bleudefrance}{rgb}{0.19, 0.55, 0.91}
\definecolor{dogwoodrose}{rgb}{0.84, 0.09, 0.41}
\hypersetup{colorlinks=true, citecolor=darkgreen, linkcolor=darkblue, 
	urlcolor = blue}

\def\be{\begin{align}}
	\def\ee{\end{align}}

\newcommand{\dd}{\mathrm{d}}
\newcommand{\DD}{\mathrm{D}}

\usepackage{newunicodechar}    
\newunicodechar{−}{\ensuremath{-}}  

\begin{document}
	
        \author{Shaofei Xu}
        \address{School of Physics and Technology, Wuhan University, Wuhan, 430072, China}

	\author{Junji Jia}
	\email[Corresponding author:~]{junjijia@whu.edu.cn}
	\address{Department of Astronomy $\&$ MOE Key Laboratory of Artificial Micro- and Nano-structures, School of Physics and Technology, Wuhan University, Wuhan, 430072, China}

    \title{Spin precession in general stationary and axisymmetric spacetimes} 
    
	\begin{abstract}
		This paper investigates the spin precession of test particles moving in the equatorial plane of general stationary and axisymmetric spacetimes using the Mathisson-Papapetrou-Dixon equations. The spin precession angles for two cases, the small-spin case and the spin-orbital plane parallel case, are derived using different approximations. For the small-spin case, the precession angle of the spin components in the equatorial plane along circular orbits is found, and perpendicular component is shown to be a constant of motion. For the spin-orbital plane parallel case, it is shown that in general the orbital and spin motions generally do not affect each other, and the spin precession angle is calculated using the post-Newtonian method to an arbitrarily high order of the orbital semi-latus rectum $p$. The precession angles in both cases are analyzed both qualitatively and quantitatively in the Kerr-Newman spacetime to elucidate their features. For large orbital radii, it is shown that the leading order of the precession angle series is generally proportional to the spacetime mass while the Lense-Thirring effect always appears from the subleading order. These precession results are then applied to various astronomical systems to determine their spin precession rates. For systems with observational data, our results show excellent agreement. For systems without observational data, we predict their spin precession rates at both the leading and Lense-Thirring effect orders. These predictions indicate that Jupiter's satellites exhibit exceptionally large geodetic spin precession and their Lense-Thirring effect may be detectable with current technology.   
	\end{abstract}
	
	\maketitle

\section{Introduction}

The motion of spinning test particle in gravitational field of a massive body is common in nature. Familiar examples include star-planet and planet-moon or satellite systems. In these systems, the test particle moves in relatively weak gravitational field where Newtonian gravitation laws apply well and general relativistic effects are typically secondary. Nevertheless, they provide natural and convenient, and sometimes unique systems for testing general relativity (GR). Systems with much stronger gravity include stars orbiting supermassive black holes (BHs) at galaxy centers, as well as binary pulsar systems. In such systems, a general relativistic treatment is often necessary to accurately describe the motion of test particles. 

Since test particles in these systems usually carry spin angular momenta, their motion will be affected by the interaction of this spin with the background spacetime provided by the more massive central objects. The motion of spinning particles in GR is typically  described by the Mathisson-Papapetrou-Dixon (MPD) equations \cite{Dixon:1964WG}. These equations have been used to study various effects, including the inspirals of spinning binaries with small mass ratios \cite{Warburton:2017sxk}, spin effects on the innermost stable circular orbit \cite{Toshmatov:2019bda} and the gravitational wave flux \cite{Skoupy:2021asz}, among others. We have previously used the MPD equations to demonstrate how a test particle's spin affects its deflection angle and gravitational lensing \cite{Zhang:2022rnn}, as well as its periapsis precession \cite{Xu:2024msu}. 

Previous solutions to the MPD equations have typically focused on either equatorial orbits or cases where the spin is perpendicular to the orbital plane. Studies in the small-spin limit have revealed that when the spin is not perpendicular to the orbital plane, spin precession occurs along the trajectory. This effect was first proposed by de Sitter, who predicted that the Earth-Moon system would exhibit a change in the longitude of the ascending node \cite{de1916einstein}. Many studies on spin precession employ either the Frenet-Serret \cite{Iyer:1993qa} or the Fermi-Walker transport frameworks \cite{straumann2013general}. Both approaches are limited by the small-spin assumption and cannot provide solutions for non-circular orbits or are restricted to specific spacetimes. The few studies used the MPD equations to investigate the spin precession-related phenomena \cite{Plyatsko:2011gf, Velandia-Heredia:2017kbc} relied on numerical methods to solve the equations, which hinder the understanding of physics and are difficult to generalize. Moreover, these studies were conducted in fixed and simple spacetimes, specifically the Schwarzschild and Kerr spacetimes. 

This paper aims to develop a systematic method to study spin precession for particles with arbitrary spin magnitudes moving along orbits that are not necessarily circular in the equatorial plane of general stationary and axisymmetric (SAS) spacetimes. 
Since these equations cannot be solved exactly to obtain the closed-form solution in the most general case, we attempt to solve them in two scenarios: the small-spin approximation for cases with a nonzero perpendicular spin component and the post-Newtonian (PN) approximation for the spin-orbital plane parallel case. These methods enable us to calculate the spin precession angles of a particle moving in the equatorial plane. The formulas for the precession angles in general SAS spacetimes are applied to the Kerr-Newman (KN) spacetime, and the effects of spacetime and orbital parameters on these angles are carefully studied. We find that the central body's spin $a$ contributes at the $p^{-3/2}$ order, where $p$ is the semi-latus rectum of the orbit, similar to the Lense-Thirring (LT) effect in orbital precession \cite{He:2023joa,Xu:2024msu}. It is also discovered that some features of the precession, namely whether the spin precession is always delayed relative to the radial direction for orbits of different radii, could be used to distinguish the naked singularity (NS) case from the BH case in the KN spacetime. 

The paper is structured as follows. In Sec. \ref{sec:msp}, we investigate the motion of spinning particles in general SAS spacetimes by analytically deriving the spin evolution equations for these particles. In Secs. \ref{sec:ssa} and \ref{sec:spinparallel}, the small-spin approximation case and spin-orbital plane parallel case are studied, respectively, to derive the corresponding spin precession formulas, and results are applied to the KN spacetime in each case. Applications of these results to relevant astronomical systems and their observations are discussed in detail in Sec. \ref{sec:app}. We conclude the paper with a discussion in Sec. \ref{sec:disc}.  Throughout the paper, the natural units $G=c=1$ and the signature convention $(-,\,+,\,+,\,+)$ are used. 

\section{Motion of spinning particles in equatorial plane
\label{sec:msp}}

We consider the motion of a spinning test particle in an SAS spacetime, whose metric in the equatorial plane can always be expressed as
\begin{align}
    \dd s^2=-g_A\dd t^2+g_{B}\dd t\dd\varphi+g_C\dd\varphi^2+g_D\dd r^2+g_F\dd\theta^2
    \label{equ:spacetime}
\end{align}
where $(t,\,\varphi,\,r,\,\theta)$ are the Boyer-Lindquist coordinates and we have set $\theta=\pi/2$ and $\dd\theta=0$ because in this work we concentrate on the motion in the equatorial plane. Consequently, the metric functions $g_A,\,g_B,\,g_C,\,g_D,\,g_F$ here are functions of $r$ only. 

The motion of a spinning particle is described by the MPD equations
\begin{align}
    \frac{\DD P^{\mu}}{\dd\tau}=&-\frac{1}{2}R^{\mu}_{\nu\rho\sigma}u^{\nu}S^{\rho\sigma},\label{equ:c4MPD1}  \\
    \frac{\DD S^{\mu\nu}}{\dd\tau}=&P^{\mu}u^{\nu}-P^{\nu}u^{\mu},\label{equ:c4MPD2}
\end{align}
where $\tau$ is the proper time and $\mathrm{D}/\dd\tau$ denotes the total derivative, $P^{\mu},\,u^{\mu},\,S^{\mu\nu}$ are the particle's four-momentum, four-velocity and antisymmetric spin tensor, respectively, and $R^{\mu}_{\nu\rho\sigma}$ is the Riemann tensor of the spacetime. 
Since Eqs. \eqref{equ:c4MPD1} and \eqref{equ:c4MPD2} contain some degeneracy, they are insufficient to solve for all unknowns, namely $P^\mu,\, u^\mu,\, S^{\mu\nu}$. 
Therefore, the following Tulczyjew condition is usually supplemented 
\begin{align}
    P_{\mu}S^{\mu\nu}=0.
    \label{equ:c4Tulczjew}
\end{align}
Additionally, we can also introduce two conserved quantities: the particle's mass $m$ and spin $J$, defined by the following equations
\begin{align}
    m^2&\equiv -P^{\mu}P_{\mu},\label{eq:mdef}\\
     J^2&\equiv \frac{1}{2}S^{\mu\nu}S_{\mu\nu}.
     \label{equ:c4mJ}
\end{align}
Furthermore, particles moving in the equatorial plane of an SAS spacetime are characterized by two conserved quantities: the particle's energy $E$ and orbital angular momentum $L$. These quantities are associated with two independent Killing vectors $\xi^{(1)}=(1,0,0,0)$ and $\xi^{(2)}=(0,0,0,1)$ through the following relations
\begin{subequations}
    \label{equ:c4Killing}
\begin{align}
    -E\equiv& P^{\mu}\xi_{\mu}^{(1)}-\frac{1}{2}S^{\mu\nu}\xi_{\mu;\nu}^{(1)},\\
    L\equiv& P^{\mu}\xi_{\mu}^{(2)}-\frac{1}{2}S^{\mu\nu}\xi_{\mu;\nu}^{(2)}.
\end{align}
\end{subequations}

Now to characterize the orientation of particle's spin, the vectorial spin $S_{\mu}$ is often defined from $S^{\mu\nu}$ and $P^\mu$ as
\begin{align}
     S_{\mu}=\frac{\sqrt{-g}}{2m}\varepsilon_{\mu\alpha\beta\gamma}S^{\alpha\beta}P^{\gamma}.
     \label{equ:c4Smu}
\end{align}
Here, $g=-g_D(g_B^2/4+g_Ag_C)$ is the determinant of the metric tensor. Once defined, it can be shown without difficulty that by inverting the linear system for $S^{\mu\nu}$ formed by Eqs. \eqref{equ:c4Tulczjew} and \eqref{equ:c4Smu} and taking advantage of the definition \eqref{eq:mdef}, that $S^{\mu\nu}$ and $S^\mu$ are reciprocal in this problem if $P^\mu$ was known, i.e., $S^{\mu\nu}$ can also be expressed as linear combinations of components of $S^\mu$ as
\begin{subequations}
\label{equ:c4Smunu}
    \begin{align}
         S^{t\varphi}=&\frac{-1}{m\sqrt{-g}}\left(g_DP^rS_{\varphi}-g_FP^{\theta}S_{r}\right),\\
        S^{tr}=&\frac{1}{2m\sqrt{-g}}\left[\left(2g_CP^{\varphi}+g_BP^t\right)S_{\theta}-2g_FP^{\theta}S_{\varphi}\right],\\
        S^{t\theta}=&\frac{1}{2m\sqrt{-g}}\left[-\left(g_BP^t+g_CP^{\varphi}\right)S_r+2g_DP^rS_{\varphi}\right],\\
        S^{\varphi r}=&\frac{1}{2m\sqrt{-g}P^{t}}\left[P^t\left(2g_AP^t-g_BP^{\varphi}\right)S_{\theta}\right.\nonumber\\
&\left.+2g_FP^{\theta}\left(P^rS_r+P^{\theta}S_{\theta}+P^{\varphi}S_{\varphi}\right)\right],\\
S^{\varphi\theta}=&\frac{1}{2m\sqrt{-g}P^t}\bigg\{2g_DP^{r}P^{\varphi}S_{\varphi}+2g_DP^rP^{\theta}S_{\theta}\nonumber\\
&+\left[2g_D\left(P^{r}\right)^2+g_BP^{\varphi}P^t-2g_A\left(P^t\right)^2\right]S_r\bigg\},\\
        S^{r\theta}=&\frac{-1}{2m\sqrt{-g}P^t}\bigg\{2\left[g_C\left(P^{\varphi}\right)^2+P^t(g_BP^{\varphi}-g_AP^t)\right]S_\varphi\nonumber\\     &+P^r(2g_CP^{\varphi}+g_BP^t)S_r+P^{\theta}\left(2g_CP^{\varphi}+g_BP^t\right)S_{\theta}\bigg\}.
    \end{align}
\end{subequations}
Note that on the right-hand sides, $S_t$ does not appear because when doing the inversion via Eq. \eqref{equ:c4Smu}, the $S_t$ equation was not used. The $S_t$ and its evolution can always be obtained once $S_i~(i=\varphi,\,r,\,\theta)$ and their evolutions are known by substituting Eq. \eqref{equ:c4Smunu} into Eq. \eqref{equ:c4Smu} for $S_t$. Therefore, henceforth most of the relations or equations of motion will not contain $S_t$. Additionally, we note that although particles moving in the equatorial plane satisfy $u^\theta=0$, the presence of particle spin means $P^\mu$ and $u^\mu$ are not parallel anymore, and thus $P^\theta$ still appears in the above equation system. 

Due to this equivalence between the tensorial and vectorial spin, we can either solve a system of equations for $(P^\mu,\, u^\mu,\, S^{\mu\nu})$ or a system for $(P^\mu,\, u^\mu,\, S^{\mu})$. However, the vectorial spin $S^\mu$ is usually easier to understand and illustrate, making it more desirable. Thus in the following, we will primarily refer to $S^\mu$ and aim to solve and discuss our results in terms of it, rather than $S^{\mu\nu}$. To this end, we convert all equations to those of $S^\mu$ first. Substituting Eq. \eqref{equ:c4Smunu} into Eqs. \eqref{equ:c4Killing}, they become
\begin{subequations}
    \label{equ:C4ENLK}
    \begin{align}
        -E=&\frac{1}{2}g_BP^{\varphi}-g_AP^t-\frac{1}{8m\sqrt{-g}}\Bigg\{2g_A^{\prime}\Big[2g_FP^{\theta}S_{\varphi}\nonumber\\
        &\left.-\left(2g_CP^{\varphi}+g_BP^t\right)S_{\theta}\Big]+g_B^{\prime}\bigg[\left(g_BP^{\varphi}-2g_AP^t\right)S_{\theta}\right.\nonumber\\
&+2g_F\frac{P^{\theta}}{P^t}\left(P^rS_r+P^{\theta}S_{\theta}+P^{\varphi}S_{\varphi}\right)\bigg]\Bigg\},\\
L=&g_CP^{\varphi}+\frac{1}{2}g_BP^t+\frac{1}{8m\sqrt{-g}}\Bigg\{g_B^{\prime}\Big[2g_FP^{\theta}S_{\varphi}\nonumber\\
        &-\left(2g_CP^{\varphi}+g_BP^t\right)S_{\theta}\Big]+2g_C^{\prime}\bigg[\left(g_BP^{\varphi}-2g_AP^t\right)S_{\theta}\nonumber\\
&+2g_F\frac{P^{\theta}}{P^t}\left(P^rS_r+P^{\theta}S_{\theta}+P^{\varphi}S_{\varphi}\right)\bigg]\Bigg\}.
    \end{align}
\end{subequations}
Here and henceforth the prime $^{\prime}$ denotes differentiation with respect to $r$.
Similarly, the $S^{\mu\nu}$ in Eq. \eqref{equ:c4Smunu} can also be substituted into definition of $J^2$. Then we recognize that in the system of equations of $P^\mu$ formed by Eqs. \eqref{eq:mdef}, \eqref{equ:c4mJ} and \eqref{equ:C4ENLK}, all of the four equations are coupled quadratic equations of $P^\mu~(\mu=t,\varphi,r,\theta)$ and therefore are not solvable analytically to obtain these components in terms of $S_\mu$ and $g_{\mu\nu}$. 
Of course, even if $P^\mu$ were analytically solvable, we still need to utilize the rest of the MPD equations to eventually solve $S_\mu$ and substitute back into $P^\mu$ so that all quantities are expressed in terms of the metric functions and proper initial conditions. 

Given this challenge, identifying special cases that admit solutions for $P^\mu$ becomes crucial. Examining Eqs. \eqref{equ:C4ENLK} again, we observe that when $P^{\theta}=0$, these equations become linear to $P^\mu$ and one can check that $P^\mu$ completely decouples from Eq. \eqref{equ:c4mJ}. Therefore under the condition that $P^\theta=0$, the system of three equations, Eqs. \eqref{equ:C4ENLK} and \eqref{eq:mdef}, becomes analytically solvable for $P^\mu~(\mu=t,\,\varphi,\,r)$. Because of this observation, next we will investigate two tractable special cases under the assumption that $P^{\theta}=0$. In Sec. \ref{sec:ssa}, the small-spin scenario, where the spin tensor (or equivalently the spin vector or $J$) is assumed small, is studied. In Sec. \ref{sec:spinparallel}, the spin-orbital plane parallel case, where the spin vector lies entirely in the equatorial plane, is analyzed. 

\section{Small-spin case\label{sec:ssa}}

\subsection{General treatment\label{subsec:smallpgp}}

For spinless particles, the four-velocity and four-momentum remain parallel, whereas this parallelism generally does not hold for spinning particles. This raises an important question: does a suitable small-spin approximation exist that preserves the parallelism between four-velocity and four-momentum?

As shown in Ref. \cite{drummond2022precisely}, the small-spin approximation provides a positive answer to this question. This approximation assumes that the spin magnitude is small, thereby allowing the neglect of higher-order spin effects in the equations of motion. Under this assumption, we first demonstrate that the evolution rate of $S^{\mu\nu}$ is second order in spin, and $P^\mu$ is proportional to $u^\mu$ to the lowest order. 

We begin by taking the total derivative on both sides of the Tulczyjew condition \eqref{equ:c4Tulczjew}
\begin{align}
        0=&\frac{\DD P_{\alpha}S^{\alpha\beta}}{\dd\tau}\nonumber\\
        =&g_{\alpha\gamma}\frac{\DD P^{\alpha}}{\dd\tau}S^{\gamma\beta}+P_{\alpha}\frac{\DD S^{\alpha\beta}}{\dd\tau}
\end{align}
where the derivative of the metric equals zero has been used. Substituting the MPD Eq. \eqref{equ:c4MPD1} into this, we derive 
\begin{align}
    0=-\frac{1}{2}g_{\alpha\gamma}R^{\alpha}_{ijk}u^{i}S^{jk}S^{\alpha\beta}+P_{\alpha}\frac{\DD S^{\alpha\beta}}{\dd\tau},
\end{align}
or equivalently
\begin{align}
    \frac{\DD S^{\mu\nu}}{\dd\tau}=\mathcal{O}(S^2)
    \label{equ:MPDlinear}
\end{align}
where $\mathcal{O}(S^n)$ stands for the $n$-th order of the tensorial/vector spin.
Substituting this into Eq. \eqref{equ:c4MPD2}, we obtain
\begin{align}
    P^{\mu}u^{\nu}-P^{\nu}u^{\mu}=\mathcal{O}(S^2).
\end{align}
Contracting with $u_{\nu}$ and using the normalization condition $u^\mu u_\mu=-1$, it becomes 
\begin{align}
    P^{\mu}=-(u_{\nu}P^{\nu})u^{\mu}+\mathcal{O}(S^2),
\end{align}
which implies that $P^{\mu}$ and $u^{\mu}$ are parallel to the leading order. 
For the proportional factor, squaring the above equation and using \eqref{equ:c4mJ} we obtain 
\begin{align}
    P^{\mu}P_{\mu}=-m^2=-(u_{\nu}P^{\nu})^2+\mathcal{O}(S^2)
\end{align}
and consequently $u_{\nu}P{\nu}= m$
and 
\begin{align}
    P^{\mu}=mu^{\mu}+\mathcal{O}(S^2).\label{eq:pup}
\end{align}
This parallelism between $P^\mu$ and $u^\mu$ under the small-spin approximation also suggests that in this case, we only need to solve one set of unknowns, $P^\mu$ or $u^\mu$. Moreover, this relation also enables us to directly set $P^\theta=0$ since $u^\theta=0$ in all of the following equations to the leading order so that many of them are greatly simplified. 

In order to find the equation of motion for the vectorial spin $S^\mu$ under the small-spin approximation, we can substitute the $S^{\mu\nu}$ in Eq. \eqref{equ:c4Smunu} into Eq. \eqref{equ:MPDlinear} and expand the system in small $S_\mu$ to first order and inverse the linear system to find $\dd S_\mu/\dd \tau$. 
In this procedure, note that according to Eq. \eqref{equ:c4MPD1} the derivative term of $P^\mu$ will not contribute to the linear order, and therefore the only question is how the $P^\mu$ themselves are expanded under the small-spin approximation. The answer to this question can be found by substituting $P^\mu$ with undetermined coefficients into Eqs. \eqref{eq:mdef} and \eqref{equ:C4ENLK} and carrying out the small $S_\mu$ expansion. This expansion yields the results 
\begin{subequations}
\label{eq:PEXP}
    \begin{align}
        P^t=&P_0^t+\frac{S_{\theta}}{m(g_B^2+4g_Ag_C)^{3/2}\sqrt{g_D}}\left(4Lg_Cg_A^{\prime}\right.\nonumber\\
        &\left.+Lg_Bg_B^{\prime}-2Eg_Cg_B^{\prime}+2Eg_Bg_C^{\prime}\right)+\mathcal{O}\left(S^{2}\right),\\
        P^{\varphi}=&P_0^{\varphi}+\frac{S_{\theta}}{m(g_B^2+4g_Ag_C)^{3/2}\sqrt{g_D}}\left(-2Lg_Bg_A^{\prime}\right.\nonumber\\
        &\left.+2Lg_Ag_B^{\prime}+Eg_Bg_B^{\prime}+4Eg_Ag_C^{\prime}\right)+\mathcal{O}\left(S^{2}\right),\\
        (P^r)^2=&(P_0^r)^2+\frac{S_{\theta}}{m(g_B^2+4g_Ag_C)^{3/2}\sqrt{g_D}}\Big[Eg_C\left(2Lg_A^{\prime}\right.\nonumber\\
        &\left.-Eg_B^{\prime}\right)-Lg_A\left(Lg_B^{\prime}+2Eg_C^{\prime}\right)+g_B\left(L^2g_A^{\prime}\right.\nonumber\\
        &\left.+E^2g_C^{\prime}\right)\Big]+\mathcal{O}\left(S^{2}\right),\\
        P^{\theta}=&\mathcal{O}\left(S^{2}\right)
    \end{align}
\end{subequations}
where the zeroth-order terms $P_0^{\mu}$, representing the four-momentum for zero spin, are given by

\begin{subequations}
\label{equ:P0}
     \begin{align}
        P_0^t=&\frac{2Lg_B+4Eg_C}{g_B^2+4g_Ag_C},\\
        P_0^{\varphi}=&\frac{4Lg_A-2Eg_B}{g_B^2+4g_Ag_C},\\
        (P_0^r)^2=&\frac{4ELg_B-m^2g_B^2+4E^2g_C-4g_A(L^2+m^2g_C)}{(g_B^2+4g_Ag_C)g_D}\label{eq:pr0eq},\\
        P_0^{\theta}=&0    .
\end{align}
\end{subequations}
Now from Eq. \eqref{equ:MPDlinear}, i.e.,
\begin{align}
    \frac{\DD}{\dd\tau}S^{\mu\nu}=\frac{\dd S^{\mu\nu}}{\dd r}u^r+(\Gamma^{\mu}_{\alpha\beta}S^{\alpha\nu}u^{\beta}+\Gamma^{\nu}_{\alpha\beta}S^{\mu\beta}u^{\alpha})=\mathcal{O}(S^2)
\end{align}
and substituting Eq. \eqref{equ:c4Smunu} and \eqref{eq:PEXP}, expanding to linear order of the spin and inverting the equation, we finally obtain 
\begin{widetext}
    \begin{subequations}
\label{equ:dSlinearS1}
    \begin{align}
    \frac{\dd}{\dd r}S_{\varphi}=&\frac{S_{\varphi}}{2}\left(\frac{Lg_B^{\prime}+2Eg_C^{\prime}}{Lg_B+2Eg_C}-\frac{g_F^{\prime}}{g_F}\right)-\frac{Sr}{2(Lg_B+2Eg_C)g_DP_0^r}\times\Big[\left(L^2+m^2g_C\right)g_B^{\prime}+\left(2EL-m^2g_B\right)g_C^{\prime}\Big]+\mathcal{O}\left(S^{2}\right),\\
        \frac{\dd}{\dd r}S_r=&\frac{2S_{\varphi}}{(Lg_B+2Eg_C)(g_B^2+4g_Ag_C)P_0^r}\Big[Eg_C(-2Lg_A^{\prime}+Eg_B^{\prime})+Lg_A(Lg_B^{\prime}+2Eg_C^{\prime})-g_B(L^2g_A^{\prime}+E^2g_C^{\prime})\Big]\nonumber\\
        &+\frac{S_{r}}{2}\left\{\frac{1}{(Lg_B+2Eg_C)(g_B^2+4g_Ag_C)}\Big[4g_C\right.\left(-2Eg_Cg_A^{\prime}+Lg_Ag_B^{\prime}\right)-4g_B\left(Lg_Cg_A^{\prime}+Eg_B^{\prime}g_C+Lg_Ag_C^{\prime}\right)\nonumber\\
        &\left.\left.+g_B^2\left(-Lg_B^{\prime}+2Eg_C^{\prime}\right)\Big]\right.\right.\left.+\frac{g_D^{\prime}}{g_D}-\frac{g_F^{\prime}}{g_F}\right\}+\mathcal{O}\left(S^{2}\right),\\
        \frac{\dd}{\dd r}S_{\theta}=&0+\mathcal{O}\left(S^{2}\right)
    \end{align}
\end{subequations}
\end{widetext}
where $P_0^r$ is given in Eq. \eqref{eq:pr0eq}.

A few comments about Eq. \eqref{equ:dSlinearS1} are in order here. We first observe from   these equations is that the spin component perpendicular to the orbital plane, $S_{\theta}$, is conserved under the linear spin approximation. In contrast, the evolution of the in-plane spin components, $S_r$ and $S_\theta$, is governed by a system of linear and homogeneous ordinary differential equations (ODE) with variable coefficients. These coefficients depend on the metric function $g_{\mu\nu}$ and the particle's kinetic variables $(E,\,L,\,m)$.
Typically, such a system does not admit closed-form solution unless the coefficient functions are very simple. Therefore to find solutions, some approximations/limits have to be taken. Second, the well-known mathematical fact that a linear ODE system with constant coefficients is solvable provides such a limit. That is, when $r$ is constant, i.e., for a circular orbit with constant radius $r=r_c$, the system of $(S_r,\,S_\varphi)$ can be solved by studying its characteristic eigen-equation. We will focus on this case in the reminder of this section.

\subsection{Spin precession along circular orbits}

To study the spin vector's circular orbit solution, we convert the derivative with respect to $r$ in Eq. \eqref{equ:dSlinearS1} to that with respect to $\varphi$ by multiplying by $\dd r/\dd\varphi=u^r/u^\varphi=P^r/P^\varphi+\mathcal{O}(S^2)=P_0^r/P_0^\varphi+\mathcal{O}(S)$, where Eq. \eqref{eq:pup} is used and $P_0^\mu$ are given in Eq. \eqref{equ:P0}. Denoting the circular orbit radius as $r_c$, the resultant equation system takes the form
\begin{subequations}
\label{equ:DSphiEL}
    \begin{align}
    \frac{\dd}{\dd\varphi}S_{\varphi}=&\frac{gS_r\Big[L(g_B^{\prime}+2Eg_C^{\prime})-m^2(g_Bg_C^{\prime}-g_B^{\prime}g_C)\Big]}{(Lg_B+2Eg_C)(2Lg_A-Eg_B)g_D^2}\Bigg|_{r=r_c}\nonumber\\
    &+\mathcal{O}\left(S^{2}\right),\\
    \frac{\dd}{\dd\varphi}S_r=&\frac{S_{\varphi}}{(Lg_B+2Eg_C)(2Lg_A-Eg_B)}\Big[L^2(g_Ag_B^{\prime}-g_A^{\prime}g_B)\nonumber\\
    &+2EL(g_Ag_C^{\prime}-g_A^{\prime}g_C)+E^2(g_Cg_B^{\prime}-g_Bg_C^{\prime})\Big]\Bigg|_{r=r_c}\nonumber\\
    &+\mathcal{O}\left(S^2\right).
\end{align}
\end{subequations}
where on the right-hand sides all functions are valuated at $r_c$. Note from this equation that for circular orbits, the coefficient matrix of $(S_r(\varphi),\, S_\varphi(\varphi))$ becomes antidiagonal, which is slightly simpler than that of Eq. \eqref{equ:dSlinearS1}.
For the circular orbit motion, the kinetic variables $(E,\, L)$ are also determined once the radius $r_c$ is fixed, due to the conditions that the radial velocity and acceleration are zero, i.e.
\begin{align}
        u^{r}|_{r=r_c}=0,\,
        \frac{\dd}{\dd r}u^{r}\Big|_{r=r_c}=0.\label{equ:ENcirular}
\end{align}
Using Eqs. \eqref{eq:pup}, \eqref{eq:PEXP} and \eqref{equ:P0}, this system can be solved under the small-spin approximation to find
\begin{subequations}
\label{equ:ELcircularre}
    \begin{align}
        \frac{E}{L}=&-\frac{\left(\frac{g_B}{g_0}\right)^{\prime}+\frac{\sqrt{(g_B^{\prime})^{2}+4g_A^{\prime}g_C^{\prime}}}{g_0}}{2\left(\frac{g_C}{g_0}\right)^{\prime}}\Bigg|_{r=r_c}+\mathcal{O}\left(S\right),\\
        \frac{1}{L}=&\frac{2\sqrt{\frac{E}{L}\left(g_B+\frac{E}{L}g_C\right)-g_A}}{m\sqrt{g_0}}\Bigg|_{r=r_c}+\mathcal{O}\left(S\right),
    \end{align}
\end{subequations}
where we have set $g_0=g_B^2+4g_Ag_C$.

Finally, substituting Eq. \eqref{equ:ELcircularre} into Eq. \eqref{equ:DSphiEL}, and solving the eigenvalues of its coefficient matrix, we can find the solution of the spin components $(S_r,\, S_\varphi)$ as 
\begin{align}
    \begin{pmatrix}
        S_{\varphi}\\
       S_r \\
    \end{pmatrix}=&
   U^{-1}
   \left(\begin{array}{cc}
        \mathrm{e}^{+i\omega_s\varphi}&0\\
        0&\mathrm{e}^{-i\omega_s\varphi}
   \end{array}\right)
    U
    \begin{pmatrix}
        S_{\varphi0}\\
        S_{r0}
    \end{pmatrix}
    \label{equ:Svarphi}
\end{align}
where $(S_{r0},\, S_{\varphi 0})$ are the initial values of the two spin components at $\varphi=0$ and the eigenvalues of the coefficient matrix are
\begin{align}
&\pm\omega_{s}=\pm\frac{\sqrt{\left(g_B^{\prime}\right)^2+4g_A^{\prime}g_C^{\prime}}}{4\sqrt{2g}g_A^{\prime}}\bigg\{g_A\left[\left(g_B^{\prime}\right)^2+4g_A^{\prime}g_C^{\prime}\right]\nonumber\\
&-\frac{g_A^{\prime}g_0^{\prime}}{2}+g_A^2\left(\frac{g_B}{g_A}\right)^{\prime}\sqrt{\left(g_B^{\prime}\right)^2+4g_A^{\prime}g_C^{\prime}}\bigg\}^{1/2}\Bigg|_{r=r_c}
    \label{equ:omegasdef}
\end{align}
and the corresponding eigen-matrix is
\begin{align}
    U=\begin{pmatrix}
        i/\sqrt{\Lambda}&-1\\
        1&-i\sqrt{\Lambda}\\
    \end{pmatrix}
\end{align}
with
    \begin{align}
    \label{eq:Clambda}
        \Lambda=&\frac{\sqrt{2g_D}}{\sqrt{g_0}g_C^{\prime}}\bigg\{g_D\Big[\left(g_B^{\prime}\right)^2+4g_A^{\prime}g_C^{\prime}\Big]-2g_C^{\prime}g_0^{\prime}\nonumber\\
        &\times(g_C^{\prime}g_B-g_Cg_B^{\prime})\sqrt{\left(g_B^{\prime}\right)^2+4g_A^{\prime}g_C^{\prime}})\bigg\}^{1/2}\Bigg|_{r=r_c}.
    \end{align}
Carrying out the matrix multiplication, solution \eqref{equ:Svarphi} can be simplified to a pure harmonic oscillation form
\begin{subequations}
\label{equ:Svarphi2}
    \begin{align}
    S_{\varphi}(\varphi)=
        &\sqrt{ S_{\varphi0}^2+S_{r0}^2/\Lambda^2}\sin\left[\omega_s\varphi+\arctan\left(\frac{\Lambda S_{\varphi0}}{S_{r0}}\right)\right],\\
S_r(\varphi)=&\sqrt{\Lambda^2 S_{\varphi0}^2+S_{r0}^2}\cos\left[\omega_s\varphi+\arctan\left(\frac{\Lambda S_{\varphi0}}{S_{r0}}\right)\right].
\label{equ:Svarphi2sr}
\end{align}
\end{subequations}
with $\omega_s$ being the oscillation frequency in $\varphi$. This $\omega_s$ then can be thought as the precession frequency of the spin.

For later comparison, we examine the large $r_c$ limit of $\omega_s$. Assuming that the metric functions have the following asymptotic expansions
\begin{subequations}
\label{eq:metricexp} 
\begin{align}
        g_A=&1+\frac{a_1}{r}+\frac{a_2}{r^2}+\dots\quad,\\
        g_B=&\frac{b_1}{r}+\frac{b_2}{r^2}+\frac{b_3}{r^3}+\dots\quad,\\
        g_C=&r^2\left(1+\frac{c_1}{r}+\frac{c_2}{r^2}\right)+\dots\quad,\\ g_D=&1+\frac{d_1}{r}+\frac{d_2}{r^2}+\dots\quad,\\
        g_F=&r^2\left(1+\frac{f_1}{r}+\frac{f_2}{r^2}+\dots\right)\quad,
    \end{align}
\end{subequations}
this frequency can be expanded in the large $r_c$ limit as 
\begin{align}
        \omega_s=&1+\frac{a_1-2d_1}{4r_c}-\frac{b_1}{2\sqrt{-2a_1}r_c^{3/2}}\nonumber\\
        &+\frac{1}{32r_c^2}\Big[-9a_1^2+4a_1(c_1-d_1)\nonumber\\
        &+4(4a_2+c_1^2-4c_2+3d_1^2-4d_2)\Big]\nonumber\\
        &-\frac{1}{4(-2a_1)^{3/2}r_c^{5/2}}\bigg\{5(a_1^2+4a_2)b_1\nonumber\\
        &+a_1\Big[-8b_2+b_1\big(c_1+2d_1\big)\Big]\bigg\}+\mathcal{O}(r_c^{-3})\label{eq:omegaslrc}
\end{align}
where the $a_i,\,b_i,\,c_i,\,d_i,\,f_i$ are the asymptotic expansion coefficients of the metric functions in Eq. \eqref{eq:metricexp}.

The $\omega_s$ actually also determines the stability of the solutions \eqref{equ:Svarphi} or \eqref{equ:Svarphi2}. When $\omega_s$ becomes imaginary, $S_r$ and $S_\varphi$ will develop exponentially growing branches and breaks the initial small-spin approximation. Studying the square root part of $\omega_s$ as in Eq. \eqref{equ:omegasdef}, we found that it contains the following factor, which should be nonnegative in order for $\omega_s$ to be real
\begin{align}
&2g_A\sqrt{\left(g_B^{\prime}\right)^2+4g_A^{\prime}g_C^{\prime}}-\bigg[g_A^2\left(\frac{g_B}{g_A}\right)^{\prime}\nonumber\\
&+\sqrt{\big[(g_Ag_B)^{\prime}\big]^2+8g_Ag_A^{\prime}(g_Ag_C)^{\prime}}\bigg]\Bigg|_{r=r_c}\geq 0
\end{align}
This inequality effectively puts some constraint on how small $r_c$ could be, i.e. $r_c$ has to be larger than some critical $r_p$, in order for the entire approach here to work. After comparing with the photon-surface (in spherical spacetimes, photon-sphere) radius that is often encountered in light-ray deflection studies, i.e., the radius below which photons incoming from infinity will not be able to escape to infinity again (see Eq. (2.18) in Ref. \cite{Duan:2023gvm}), we found that $r_p$ is exactly the photon sphere radius there. Since $r_p$ for typical space times is small (e.g. $3M$ for Schwarzschild spacetime), we expect that the analysis here is valid for very large range of the circular orbit radius.

On the other hand, when the object moves along a circular geodesic, we can work out its four-velocity to the $\mathcal{O}(S^0)$ order by substituting the energy and angular momentum in Eq. \eqref{equ:ELcircularre} into Eq. \eqref{equ:P0} and using the relation \eqref{eq:pup}. The result is found to be
\begin{subequations}
    \begin{align}
        u^t_0=&1/\sqrt{g_A-\omega_o g_B-\omega_o^2g_C}\Big|_{r=r_c},\\
u^{\varphi}_0=&\omega_o u^{t}_0,\\
u^r_0=&u^{\theta}_0=0.
    \end{align}
\end{subequations}
with the orbital rotation frequency 
\begin{align}       
\omega_o=&\frac{-g_B^{\prime}+\sqrt{\left(g_B^{\prime}\right)^2+4g_A^{\prime}g_C^{\prime}}}{2g_C^{\prime}}\Bigg|_{r=r_c}
\end{align} 
Using this orbital trajectory solution, we can define an observer that is comoving with the spining object, and its local tetrad becomes {\cite{Seme:1993} 
\begin{subequations}
\label{eq:tetrad}
    \begin{align}
        e^{\mu}_{(\hat{t})}=&(u^t_0,\,u^t_0\omega_o,\,0,\,0)\equiv Z^\mu,\\
e^{\mu}_{(\hat{\varphi})}=&\frac{(g_B+2\omega_o g_C,\,2g_A-\omega_o g_B,\,0,\,0)}{\sqrt{g_0(g_A-\omega_o g_B-\omega_o
        ^2g_C)}},\\
        e^{\mu}_{(\hat{r})}=&(0,\,0,\,\frac{1}{\sqrt{g_D}},\,0),\\
        e^{\mu}_{(\hat{\theta})}=&(0,\,0,\,0,\,\frac{1}{\sqrt{g_F}}).
    \end{align}
\end{subequations}
where $Z^\mu$ is the four-velocity of the observer and $(\hat{t},\,\hat{\varphi},\,\hat{r},\,\hat{\theta})$ are the coordinates of the tetrad. The spin tensor $S^{\mu\nu}$ and momentum vector $P^\mu$ in the old coordinates can be transformed into $S^{\hat{\mu}\hat{\nu}}$ and $P^{\hat{\mu}}$ in the new frame using the tetrad \eqref{eq:tetrad}. Here $\hat{\mu},\,\hat{\nu}$ etc. run through $(\hat{t},\,\hat{\varphi},\,\hat{r},\,\hat{\theta})$. Then the new spin vector $S_{\hat{\mu}}$ can be related, using definition 
\begin{align}
     S_{\hat{\mu}}=\frac{\sqrt{-\hat{g}}}{2m}\varepsilon_{\hat{\mu}\hat{\alpha}\hat{\beta}\hat{\gamma}}S^{\hat{\alpha}\hat{\beta}}P^{\hat{\gamma}}.
     \label{equ:SmuE4}
\end{align} where $\hat{g}=-1$ now, to the old one by
    \begin{align}
    \label{eq:sphat}
S_{\hat{\mu}}=\left(0,\,2\sqrt{\frac{g_F}{g_0}}\frac{S_{\varphi}}{u^t_0},\,\sqrt{\frac{g_F}{g_D}}S_r,\,S_{\theta}\right)\bigg|_{r=r_c}
    \end{align}
where $S_r$ and $S_\varphi$ are given in Eq. \eqref{equ:Svarphi2}. One can easily check that this spin vector satisfies the normalization condition 
\begin{align}
S_{\hat{\mu}}S^{\hat{\mu}}=&S_{\hat{\varphi}}^2+S_{\hat{r}}^2+S_{\hat{\theta}}^2\nonumber\\
    =&\frac{4g_F}{(u^t_0)^2g_0}S_{\varphi}^2+\frac{g_F}{g_D}S_r^2+S_{\theta}^2\bigg|_{r=r_c}=J^2
\end{align}
where the last equality can be established by substituting solution \eqref{equ:Svarphi2} into Eq. \eqref{equ:c4Smu} and further into condition \eqref{equ:c4mJ}.

The spin vector \eqref{eq:sphat} clearly is spacelike and its two components within the equatorial plane actually varies along the orbit with the azimuth angle $\varphi$.
A meaningful quantity for the local observer to measure is the angle that these two components $(S_{\hat{\varphi}},\, S_{\hat{r}})$ make against a fixed direction in the local tetrad, which we choose to be the radial $\hat{r}$ direction. Using the cosine theorem and Eqs. \eqref{equ:Svarphi2} and \eqref{eq:sphat}, the angle from $\hat{r}$ to the spin direction can be computed as
\begin{align}
    \alpha_{\hat{r}\to\hat{S}}=&- \arccos\frac{S_{\hat{r}}}{|S_{\hat{\phi}}^2+S_{\hat{r}}^2|^{1/2}}\bigg|_{r=r_c}\nonumber\\
    =&-\omega_s\varphi-\arctan\left(\frac{\Lambda S_{\varphi0}}{S_{r0}}\right)\label{eq:alphacirc}
\end{align}
where $\omega_s$ and $\Lambda$ were given in Eqs. \eqref{equ:omegasdef} and \eqref{eq:Clambda}. The $S_{\varphi 0}$ and $S_{r0}$ are determined by the initial spin orientation. If the observer were measuring the precession of the spin direction against a remote fixed celestial body along the $\hat{x}^+$ axis, then the precession angle $\alpha_{\hat{x}^+ \to\hat{S}}$ should equal the $\alpha_{\hat{r}\to \hat{S}}$ should be compensated by $\varphi$, i.e.,
\begin{align}
    \alpha_{\hat{x}^+ \to\hat{S}}=(1-\omega_s)\varphi-\arctan\left(\frac{\Lambda S_{\varphi0}}{S_{r0}}\right) .\label{eq:alphacirc2}
\end{align}

Eq. \eqref{eq:alphacirc} and \eqref{eq:alphacirc2} mean that the spin measured by the observer evolves as $\varphi$ increases along the circular orbit with a constant angular velocity $\omega_s$. To quantitatively analyze this precession characterized by $\omega_s$, next we study these quantities in the KN spacetime.

\subsection{Spin precession along circular orbits in KN spacetime}

In this subsection, we apply the results obtained above to the evolution of the small spin along circular orbit in the KN spacetime, whose metric functions in the form of Eq. \eqref{equ:spacetime} are
\begin{subequations}
\label{eq:knm}
\begin{align}
        g_A(r)=&1-\frac{2Mr-Q^2}{r^2},\\
        g_B(r)=&-\frac{2a (2Mr-Q^2)}{r^2},\\
        g_C(r)=&r^2+a ^2+\frac{2a ^2Mr-a ^2Q^2}{r^2},\\
        g_D(r)=&\frac{r^2}{\Delta},\,\Delta\equiv r^2-2Mr+a ^2+Q^2,\\
        g_F(r)=&r^2
    \end{align}
\end{subequations}
where $M,\,Q$ and $a$ are the mass, charge and angular momentum per unit mass of the spacetime respectively.
Using Eq. \eqref{equ:omegasdef}, the spin precession frequency $w_s$ in this case can be obtained as
\begin{align}
\omega_s=\sqrt{1-\frac{3M}{r_c}+\frac{Q^2+2a \sqrt{Mr_c-Q^2}}{r_c^2}}. \label{eq:omegaskn}
\end{align}
The corresponding value of this precession frequency in the Schwarzschild spacetime, i.e. after setting $a=Q=0$, was also obtained in Eq. (4) of Ref. \cite{dolan2014gravitational} and Eq. (50) of Ref. \cite{chakraborty2017distinguishing}.

Several points about the frequency \eqref{eq:omegaskn} are worth mentioning here. First, we note that when $r_c$ is much larger than $M,\,Q$ and $a$, i.e., in the weak field regime, $\omega_s$ approaches 1 from below. Indeed, we can show that this upper bound of 1 is true for all physical $r_c$ outside the outer event horizon of the KN BH. 
Expanding \eqref{eq:omegaskn} in the large $r_c$ limit and using Eq. \eqref{eq:alphacirc2}, we see that the frequency $\omega_s$ and the spin direction $\alpha_{\hat{x}^+ \to\hat{S}}$ to the leading three orders become
\begin{align}
\omega_s= 1-\frac{3M}{2r_c}&+\frac{a\sqrt{M}}{r_c^{3/2}}-\frac{9M^2-4Q^2}{r_c^2}+\mathcal{O}(r_c^{-5/2}),\label{eq:omegasasy}\\
\alpha_{\hat{x}^+ \to\hat{S},\mathrm{KN}}=&\left(\frac{3M}{2r_c}-\frac{a\sqrt{M}}{r_c^{3/2}}+\frac{9M^2-4Q^2}{r_c^2}\right)\varphi\nonumber\\
&-\arctan\left(\frac{\Lambda S_{\varphi0}}{S_{r0}}\right) .
\label{eq:spindirinkn}
\end{align}
It is clear then at the leading non-trivial order, the precession frequency decreases in the inverse power of the circular orbit radius. The spin $a$ enhances (or decreases) the evolution when its corotating (or counter-rotating) with the orbit at half an order of $r_c$ higher. While the effect of spacetime charge $Q$ is at even higher order, increasing the frequency regardless of its sign. 

The second point is about the behavior of $\omega_s$ when $r_c$ is small, i.e., near the ergosphere when the spacetime is a BH one or near the NS. For the Kerr case with $Q=0$, it was known in Ref. \cite{chakraborty2017distinguishing} that for the NS case with $|a|>M$ the precession frequency diverges as $r_c$ approaches the ring singularity at $r=0,\,\theta=\pi/2$,  but only takes a finite value when approaching the event horizon from the equatorial plane for a BH one. For the KN spacetime, we can show that this is not the case any longer. When $Q\neq 0$ for the NS case, even when $r$ approaches the minimal radius allowed for circular orbits, $\omega_s$ will not approach infinity but zero. For example, when $Q=M/2,\,a=1951M/1024$, it is not difficult to verify that the minimal circular orbit radius is $r_c=17M/64$, which when approached generates $\omega_s=0$. 

Since in the KN case the divergence/convergence of the precession frequency can not be used as a solid criteria to distinguish BH and NS spacetimes, we propose to use whether $\omega_s$ can be larger than 1 to distinguish the KN BH and NS cases.
What we can show here is that 
for a KN BH spacetime, $\omega_s$ will have a upper limit of 1. In other words, the spin direction always precesses slower than the radial direction. This limit can be shown as the following. Since $(|a|,\,|Q|)\leq M$ for KN BH, it is evident that in Eq. \eqref{eq:omegaskn}, the term $(Q^2+2a \sqrt{Mr_c-Q^2})/r_c^2\leq (M^2+2M \sqrt{Mr_c})/r_c^2$, and therefore Eq. \eqref{eq:omegaskn} can yield
\begin{align} 
\omega_s&\leq\sqrt{1-\frac{3M}{r_c}+\frac{M^2+2M \sqrt{Mr_c}}{r_c^2}}\nonumber\\
&=\sqrt{1-(\sqrt{r_c}-\sqrt{M})(3\sqrt{r_c}+\sqrt{M})\frac{M}{r_c^2}}. \end{align} 
Now because $r_c>r_{\mathrm{H}+}=M+\sqrt{M^2-Q^2}>M$ in the equatorial plane, we see immediately that $\omega_s<1$. 
For the NS case however, one can easily see that in the large $a$ limit, Eq. \eqref{eq:omegaskn} can always reach a value larger than 1.  
Therefore, the above proof and example shows that for the KN spacetime, whether $\omega_s$ can reach a value larger than $1$ corresponds to the BH and NS situations very well and can be used to distinguish these cases. 

\begin{figure}[htp!]
\centering
\includegraphics[width=0.48\textwidth]{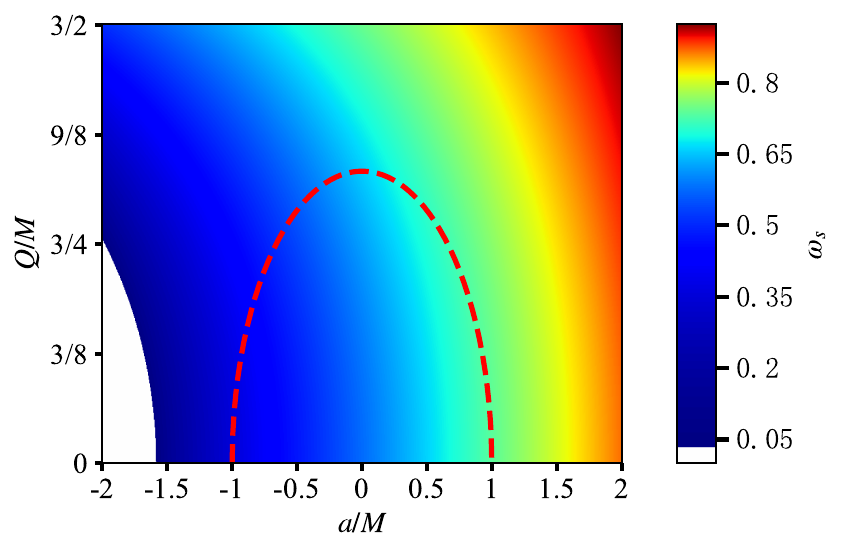}\\
(a)\\
\includegraphics[width=0.48\textwidth]{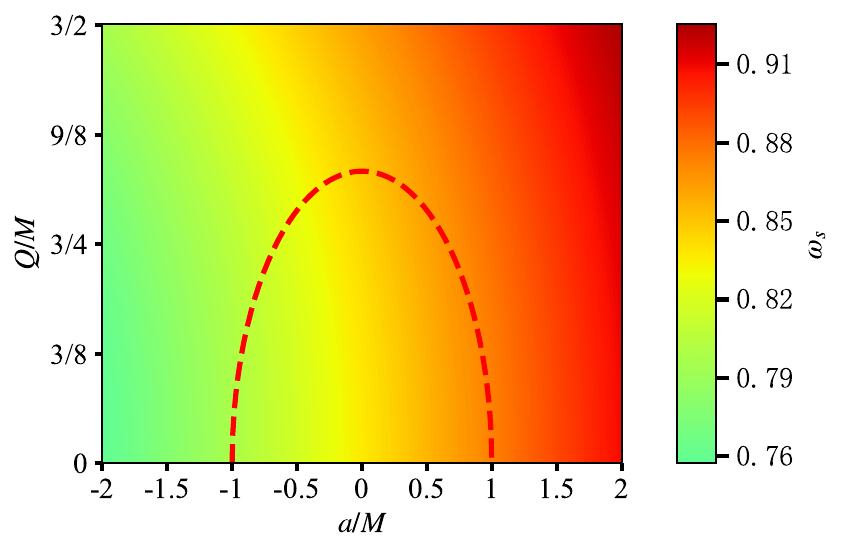}\\
(b)
\caption{Spin precession frequency $\omega_s$ as function of $a,\, Q$ for different $r_c$: (a) $r_c=9M/2$, (b) $r_c=10M$ respectively. 
The red dashed line represents the BH/NS boundary in the parameter space. The white region in (a) indicates where $\omega_s$ becomes imaginary or equivalently no circular orbit of that size exists.}
  \label{fig:omegas}
\end{figure}

In Fig. \ref{fig:omegas}, we plot the $\omega_s$ in Eq. \eqref{eq:omegaskn} for different values of the spacetime spin $a $ and  charge $Q$ for several $r_c$. 
Note that since the precession of the spin does not require the spacetime to be a BH one, we allow the parameters $a$ and $Q$ to vary beyond their BH limits so that the BH can turn into a NS. 
It is seen by comparing plots (a) and (b) for points with same values of $(a,\,Q)$ that $\omega_s$ increases universally for all $a$ and $Q$ as $r_c$ increases. Moreover, in both plots, we observe that $\omega_s$ increases with the increase of both $a$ and $Q$ but much stronger with the former. These all agree with the analysis done below Eq. 
\eqref{eq:omegasasy} although $r_c$ here is not asymptotically large yet. In addition, we also note that the $\omega_s$ transforms smoothly when the spacetime goes from a BH one to a NS or vice versa. Although in the NS region in both plots of Fig. \ref{fig:omegas} the $\omega_s$ is still not larger than 1, this will quickly change if we further enlarge the parameter space (especially when $|a|$ is very large).

\begin{figure}[htp!]
\begin{centering}
\includegraphics[width=0.48\textwidth]{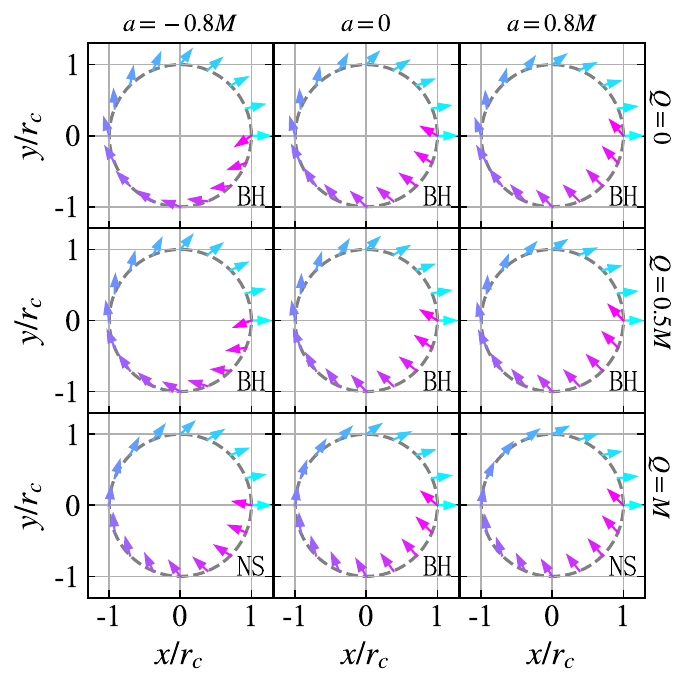}\\
(a)\\
\includegraphics[width=0.48\textwidth]{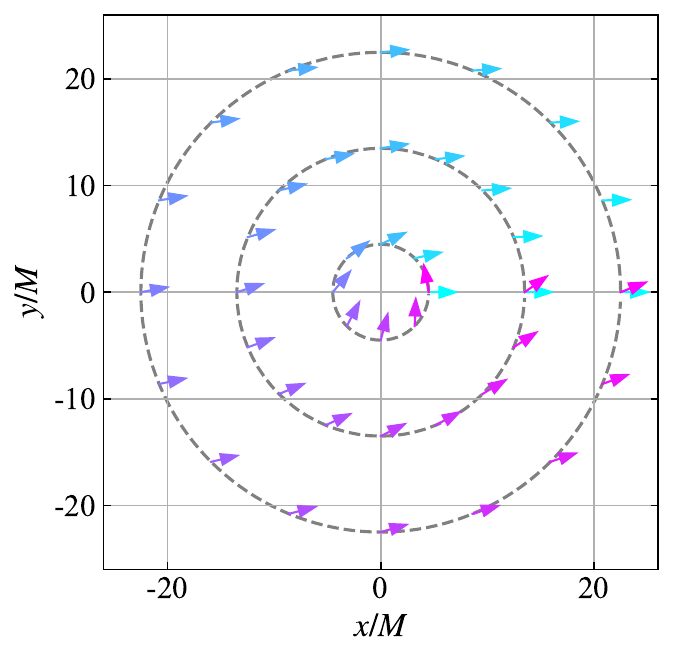}\\
(b)
\end{centering}
\caption{(a) Spin precession along circular orbits for different $a$ and $Q$. $a=-0.9M,\,0,\,0.9M$ respectively in the first to third columns and $Q=0,\,0.5M,\,M$ respectively in the bottom to top rows. $r_c=9M/2$ is set in this plot.  (b) Spin precession for different radius $r_c$. From inner to outer: $r_c=9M/2,\,27M/2,\,45M/2$. $Q=M/2$ and $a=4M/5$ are used in this plot. $S_{\varphi 0}=0$ is chosen as initial condition at $\varphi=0$ in both plots. Arrows denoting the spin orientations are color coded linearly with the precession angle. BH and NS in the corners indicate the corresponding spacetimes.}
\label{fig:imagedS}
\end{figure}

Finally, in Fig. \ref{fig:imagedS} we present the precession of the particle's spin components $(S_{\hat{r}},\, S_{\hat{\varphi}})$ in the equatorial plane during its motion along circular orbits, as measured by the comoving observer defined in Eq. \eqref{eq:tetrad}. The $S_{\hat{\theta}}$ component is not drawn because it is a constant of motion. In Plot (a), the spin precession angle against the radial direction is plotted according to Eq. \eqref{eq:alphacirc} for several $(a,\,Q)$ combinations picked from Fig. \ref{fig:omegas} (a), including both the BH and NS cases, as indicated in the corner of each grid. The $S_{\varphi 0}=0$ and $S_{r0}>0$ initial condition is chosen at $\varphi=0$, i.e., the spin points along the $\hat{x}^+$ direction initially. It is seen that the angle from the spin to $\hat{r}$ accumulates negatively and linearly as $\varphi$ increases, due to the fact that $0<\omega_s<1$ for the parameters we choose. For the effects of $a$ and $Q$ on the speed of precession, we see that they agree with what was observed in Fig. \ref{fig:omegas}. That is, the larger the $a$ and $Q$, the faster the spin precesses, and the former influences the precession stronger. Note that in plot (a), a small $r_c=9M/2$ is chosen in order for the precession to be noticeable by eyes. In plot (b) however, we fix $(Q,\,a)$ but chose three linearly increasing $r_c$. It is seen that as $r_c$ increases, then the precession against the $\hat{x}^+$ direction decreases monotonically. If an even larger $r_c$ were used, the spin direction would just tilt up very slightly as the object rotates a full circle, as dictated by Eq. \eqref{eq:spindirinkn}. 

\section{Spin-orbital plane parallel case\label{sec:spinparallel}}

\subsection{Geneal treatment}

As discussed in Sec. \ref{sec:msp} and observed in Eqs. \eqref{eq:mdef}, \eqref{equ:c4mJ} and \eqref{equ:C4ENLK}, analytically solving the motion of particles with arbitrary spin orientations without any spin approximation is challenging. 
In this section, we consider a special case that the vectorial spin lies in the equatorial plane, i.e., the orbital plane. It turns out that in this case, the orbital motion and spin evolution will not affect each other, allowing the spin vector to be solved analytically under the PN assumption. 

We begin by making an observation that if we assume 
\begin{align}
\label{eq:pthetastheta0}    P^{\theta}=0=S_{\theta}
\end{align} 
in Eq. \eqref{eq:mdef} and \eqref{equ:C4ENLK}, these equations completely decouple from $S_\mu~(\mu=t,\,r,\,\theta,\,\varphi)$, indicating that the four velocity $P^\mu$, or equivalently the orbital motion, is not affected by the spin of the particle at all in this scenario. 
Thus, in this subsection, we will concentrate on the case $S_{\theta}=0$, i.e., the particle's spin lies entirely within the orbital plane. And we also assume that $P^{\theta}\equiv0$, which is required by motion in the equatorial plane.

Under the assumption \eqref{eq:pthetastheta0}, Eq. \eqref{equ:c4Smunu} for  $S^{\mu\nu}$ reduces to
\begin{subequations}
    \label{equ:Sijpar}
    \begin{align}
    S^{tr}=&S^{t\varphi}=S^{r\varphi}=0,\\
    S^{t\theta}=&\frac{g_DP^rS_{\varphi}-(g_CP^{\varphi}+g_BP^t/2)S_r}{m\sqrt{-g}},\\
    S^{r\theta}=&\frac{1}{m\sqrt{-g}P^t}\Big\{\left[g_A(P^t)^2-g_BP^{\varphi}P^t-g_C(P^{\varphi})^2\right]S_{\varphi}\nonumber\\
    &-(g_CP^{\varphi}+g_BP^t/2)P^rS_r\Big\},\\
    S^{\varphi\theta}=&\frac{P^r}{m\sqrt{-g}P^t}\Big\{g_DP^{\varphi}S_{\varphi}\nonumber\\
    &-\left[g_A(P^t)^2-g_BP^{\varphi}P^t/2-g_D(P^{r})^2\right]S_r\Big\}.
    \end{align}
\end{subequations}
The four-momentum here can be solved from Eqs. \eqref{equ:C4ENLK} and \eqref{eq:mdef} under the assumption \eqref{eq:pthetastheta0} too, to obtain
\begin{subequations}
\label{equ:Ppar}
    \begin{align}
        P^t=&\frac{2Lg_B+4Eg_C}{g_B^2+4g_Ag_C},\\
        P^{\varphi}=&\frac{4Lg_A-2Eg_B}{g_B^2+4g_Ag_C},\\
        (P^r)^2=&\frac{4ELg_B-m^2g_B^2+4E^2g_C-4g_A(L^2+m^2g_C)}{(g_B^2+4g_Ag_C)g_D},\label{equ:Pparr}\\
        P^{\theta}=&0.
    \end{align}
\end{subequations}
Substituting Eq. \eqref{equ:Sijpar} and Eq. \eqref{equ:Ppar} into the right and left-hand sides of the MPD Eq. \eqref{equ:c4MPD1} respectively, after some simplification we can find that the four-velocity takes the form 
\begin{align}
\label{equ:upar}
    u^{\mu}=P^{\mu}/m.
\end{align}
The absence of the spin vector or tensor in this equation and Eq. \eqref{equ:Ppar} demonstrates that the test particle's motion depends solely on the metric components $g_{\mu\nu}$ and the particle's energy $E$ and angular momentum $L$, as if no spin were present, confirming our earlier claim that under condition \eqref{eq:pthetastheta0}, the particle's spin does not affect its orbital motion. 

Finally, regarding the evolution of the particle's spin, we can substitute Eq. \eqref{equ:Sijpar} and  \eqref{equ:Ppar} into the MPD Eq. \eqref{equ:c4MPD2} and obtain 
\begin{widetext}
    \begin{subequations}
\label{equ:dSpar}
    \begin{align}
        \frac{\dd}{\dd r}S_r=&\frac{2S_{\varphi}}{(Lg_B+2Eg_C)(g_B^2+4g_Ag_C)P^r}\Big[Eg_C(-2Lg_A^{\prime}+Eg_B^{\prime})+Lg_A(Lg_B^{\prime}+2Eg_C^{\prime})-g_B(L^2g_A^{\prime}+E^2g_C^{\prime})\Big]\nonumber\\
        &+\frac{S_{r}}{2}\left\{\frac{1}{(Lg_B+2Eg_C)(g_B^2+4g_Ag_C)}\Big[4g_C\right.\left(-2Eg_Cg_A^{\prime}+Lg_Ag_B^{\prime}\right)-4g_B\left(Lg_Cg_A^{\prime}\right.\nonumber\\
        &\left.\left.\left.+Eg_B^{\prime}g_C+Lg_Ag_C^{\prime}\right)+g_B^2\left(-Lg_B^{\prime}+2Eg_C^{\prime}\right)\Big]\right.\right.\left.+\frac{g_D^{\prime}}{g_D}-\frac{g_F^{\prime}}{g_F}\right\}\\
        \frac{\dd}{\dd r}S_{\varphi}=&\frac{S_{\varphi}}{2}\left(\frac{Lg_B^{\prime}+2Eg_C^{\prime}}{Lg_B+2Eg_C}-\frac{g_F^{\prime}}{g_F}\right)-\frac{Sr}{2(Lg_B+2Eg_C)g_DP^r}\Big[\left(L^2+m^2g_C\right)g_B^{\prime}+\left(2EL-m^2g_B\right)g_C^{\prime}\Big].
    \end{align}
\end{subequations}
\end{widetext}
This is a system of coupled and homogeneous linear first order ODEs with variable coefficients of $S_r(r)$ and $S_\varphi(r)$. It turns out that they can be decoupled via a linear transformation into two separate second-order ODEs, as shown in Eq. \eqref{equ:dhatS2} of Appendix \ref{app:eqsol}. However, these new equations still have variable coefficients, and thus no closed-form solutions can be obtained immediately for general metric. To solve them, we will have to use further assumptions that can simplify the system to some solvable form. In next subsection, we will employ the PN method to these equations. That is, to study their solutions when the orbital radius is much larger than the mass of spacetime. 

\subsection{PN method and solution}

In this subsection, we will employ the PN approximation to solve the spin evolution equations \eqref{equ:dhatS2} in the weak-gravity limit and calculate the spin precession in the orbital plane. The PN approximation was used as an asymptotic approach in our previous work Ref. \cite{He:2023joa} to calculate the periapsis precession angle. 

We begin by noting that in Newtonian gravity, the orbit equation for bound orbit is
\begin{align}
    \frac{1}{r}=\frac{1+e\cos\varphi}{p},
\end{align}
where $p$ is the semi-latus rectum, $e$ is the eccentricity.
The PN approximation assumes that under weak field conditions, the particle trajectories in GR can be described as
\begin{align}
    \frac{1}{r}=\frac{1+e\cos\psi(\varphi)}{p}
    \label{equ:pnorbit}
\end{align}
Here $\psi$ is an unknown function of $\varphi$. When $p$ is large (the Newtonian limit), it is expected that $\psi$ is a function that deviates not too fast from $\varphi$. Moreover, without losing any generality we can choose $\psi|_{\varphi=0}=0$. 
Clearly, when $\psi=0$ or $\psi=\pi$, the orbit reaches its apoapsis $r_+$ or periapsis $r_-$ respectively, i.e., 
\begin{align} \label{eq:r1r2def}
    \frac{1}{r_+}=\frac{1+e}{p},\quad     \frac{1}{r_-}=\frac{1-e}{p}. 
\end{align}

Under the PN approximation, we can solve the spin procession equations \eqref{equ:dSpar} in the large $p$ limit. Since the relation \eqref{equ:pnorbit} provides a one to one mapping from the new angle variable $\psi$ to the radius $r$, we will express $S_r$ and $S_\varphi$ using variable $\psi$. The solution process is slightly tedious and therefore only shown in Appendix \ref{app:eqsol}, and the results are
\begin{widetext}
\begin{subequations}
\label{equ:srsphisol}
\begin{align}
    S_r(\psi)=&\Bigg\{\frac{(1+e\cos\psi)\cos(\psi+\eta)}{1+e}+\frac{e(1+e\cos\psi)}{(1+e)p}\Big[2c_1\cos\eta(1-\cos\psi)+(2d_1+ea_1-2f_1)\sin\eta\sin\psi\nonumber\\
    &-2(c_1-d_1+f_1)\cos\psi\cos(\psi+\eta)\Big]-\frac{eb_1(1+e\cos\psi)\sin\psi\sin(\psi+\eta)}{\sqrt{-8a_1}(1+e)p^{3/2}}+\mathcal{O}(p^{-2})\Bigg\}S_{r0}\nonumber\\
    &+\left[\frac{(1+e\cos\psi)\sin(\psi+\eta)}{p}+\mathcal{O}(p^{-3/2})\right]S_{\varphi0},\label{eq:srpsisol}\\
    S_\varphi(\psi)=&\Bigg\{\cos(\psi+\eta)+\frac{e}{4p}\left[\left(2(a_1-c_1)-2f_1\cos\psi\right)\cos(\psi+\eta)-2(a_1-c_1)\cos\eta-(ea_1+2f_1)\sin\psi\sin\eta\right]\nonumber\\
    &-\frac{eb_1\sin\psi\sin(\psi+\eta)}{\sqrt{-8a_1}p^{3/2}}+\mathcal{O}(p^{-2})\Bigg\}S_{\varphi0}-\Bigg\{\frac{\sin(\psi+\eta)p}{1+e}-\frac{1}{4(1+e)a_1}\bigg\{2ea_1(-a_1+c_1)\sin(\psi-\eta)\nonumber\\
    &+2\Big[8a_2+a_1\big((-7+e)a_1-(-6+e)c_1-2(1+e)d_1+2ef_1\big)\Big]\sin(\psi+\eta)+ea_1\Big[(4+e)a_1-4c_1\Big]\sin\eta\nonumber\\ &+2ea_1f_1\cos\psi\sin(\psi+\eta)\bigg\}+\frac{b_1[e\cos\psi\sin\eta+(6-e)\sin(\psi+\eta)]}{(1+e)\sqrt{-8a_1p}}+\mathcal{O}(p^{-1})\Bigg\}S_{r0}
\end{align}
\end{subequations}
\end{widetext}
where 
\begin{align}
\label{equ:dualscale}
\eta=\sum_{i=2}\frac{\omega_i}{p^{i/2}}\psi
\end{align}
with its first few coefficients
\begin{subequations}
\label{eq:omegaires}
    \begin{align}
        \omega_2=&-\frac{3a_1^2-4a_2-2a_1c_1}{4a_1},\label{eq:omega2res}\\
        \omega_3=&-\frac{3b_1}{\sqrt{-8a_1}},\label{eq:omega3res}\\
        \omega_4=&-\frac{1}{32a_1^2}\Big[\left(6e^2-63\right)a_1^4+(56c_1-16e^2)a_1^3\nonumber\\
        &+\left(8e^2c_1^2-48c_2-8e^2a_2+136a_2\right)a_1^2\nonumber\\
        &+\left(-96a_3+16a_2c_1-64a_2c_1\right)a_1+16a_2^2\Big].
    \end{align}
\end{subequations}
and the $a_i,\,b_i,\,c_i,\,d_i,\,f_i$ are the coefficients of the metric functions in Eq. \eqref{eq:metricexp}. 

\subsection{Precession between two apoapsides}

Since in the PN approximation, the orbit is not necessarily a circle, from the observational point of view, one might be more interested in the precession of the spin angle when the object passes through a fixed radius. In this subsection, we will study the precession angle of the spin vector after the spin returns to its apoapsis (or periapsis) each time, as observed by an static observer at the apoapsis (or periapsis).

The tetrad for an static observer in the equatorial plane of this spacetime is given by 
\begin{subequations}
\label{eq:tetradso}
    \begin{align}
        e^{\mu}_{(\bar{t})}=&(\frac{1}{\sqrt{g_A}},0,0,0)\\
        e^{\mu}_{(\bar{\varphi})}=&(\frac{g_B}{\sqrt{g_Ag_0}},2\sqrt{\frac{g_A}{g_0}},0,0)\\
        e^{\mu}_{(\bar{r})}=&(0,0,\frac{1}{\sqrt{g_D}},0)\\
        e^{\mu}_{(\bar{\theta})}=&(0,0,0,\frac{1}{\sqrt{g_F}})
    \end{align}
\end{subequations}
The spin tensor $S^{\mu\nu}$ and momentum vector $P^\mu$ in the coordinates $(t,\,\varphi,\,r,\,\theta)$ can be transformed into $S^{\bar{\mu}\bar{\nu}}$ and $P^{\bar{\mu}}$ in the new frame $(\bar{t},\,\bar{\varphi},\,\bar{r},\,\bar{\theta})$ using the tetrad \eqref{eq:tetradso}. Then the new spin vector in the frame of the observer can be found, using definition
\begin{align}
     S_{\bar{\mu}}=\frac{\sqrt{-\bar{g}}}{2m}\varepsilon_{\bar{\mu}\bar{\alpha}\bar{\beta}\bar{\gamma}}S^{\bar{\alpha}\bar{\beta}}P^{\bar{\gamma}}
     \label{equ:SmuE4}
\end{align}
to be

\begin{subequations}
\label{eq:smubar}
    \begin{align}        S_{\bar{t}}=&-\frac{\sqrt{g_F}(P^{\varphi}S_{\varphi}+P^rS_r)}{\sqrt{g_A}P^t}\\    S_{\bar{\varphi}}=&\frac{\sqrt{g_F}\Big[2g_AP^tS_{\varphi}-g_B\big(P^{\varphi}S_{\varphi}+P^rS_r\big)\Big]}{\sqrt{g_Ag_0}P^t}\\    S_{\bar{r}}=&\frac{\sqrt{g_F}S_r}{\sqrt{g_D}}\\        S_{\bar{\theta}}=&0    \end{align}
\end{subequations}
where the spin vector $S_r$ and $S_\varphi$ are given in Eq. \eqref{equ:srsphisol} and $P^\mu$ are in Eq. \eqref{equ:Ppar}.

To determine the precession angle of the spin vector from one apoapsis to the next, we substitute $\psi=0$ and $\psi=2\pi$ respectively into Eq. \eqref{eq:smubar} to obtain 
\begin{widetext}
    \begin{subequations}
\label{equ:SbrieSr0}
    \begin{align}
    S_{\bar{r}}|_{\psi=0}=&\Big[\frac{p}{1+e}+\frac{f_1-d_1}{2}+\frac{(1+e)(3d_1^2-4d_2-2d_1f_1-f_1^2+4f_2)}{8p}+\mathcal{O}(p^{-2})\Big]S_{r0}\\
S_{\bar{\varphi}}|_{\psi=0}=&\Big[1-\frac{(1+e)(a_1-c_1-d_1+f_1)}{p}-\frac{(1+e)^2}{8p^2}\Big[(a_1+c_1+d_1-f_1)^2-4(c_1-f_1)(c_1+d_1)\nonumber\\
&-4(a_2-c_2+d_1^2+f_2)\Big]+\mathcal{O}(p^{-5/2})\Big]S_{\varphi0}\\
S_{\bar{r}}|_{\psi=2\pi}=&\Bigg\{\frac{p}{1+e}+\frac{f_1-d_1}{2}+\frac{1}{8a_1^2(1+e)p}\bigg\{(1+e)^2a_1^2\Big[-4d_2+(d_1-f_1)(3d_1+f_1)+4f_2\Big]\nonumber\\
&+(3a_1^2-4a_2-2a_1c_1)^2\pi^2\bigg\}+\frac{3\pi^2(3a_1^2-4a_2)b_1}{(1+e)(-2a_1p)^{3/2}}+\mathcal{O}(p^{-2})\Bigg\}S_{r0}\nonumber\\
&-\Big[\frac{\pi(3a_1^2-a_1c_1-2a_2)}{2a_1P}+\frac{3\pi b_1}{\sqrt{-2a_1}p^{3/2}}+\mathcal{O}(p^{-2})\Big]S_{\varphi0}\\
S_{\bar{\varphi}}|_{\psi=2\pi}=&\Bigg\{1-\frac{(1+e)\Big[(1+e)a_1+2(c_1+d_1-f_1)\Big]}{4p}+\mathcal{O}(p^{-2})\Bigg\}S_{\varphi0}\nonumber\\
&+\frac{\pi}{1+e}\Big[\frac{(3a_1^2-4a_2-2a_1c_1)}{2a_1}+\frac{3 b_1}{\sqrt{-2a_1p}}+\mathcal{O}(p^{-1})\Big]S_{r0}\\
S_{\bar{t}}|_{\psi=0,2\pi}=&-\frac{\sqrt{g_0g_Ag_F}P^tS_{\bar{\varphi}}|_{\psi=0,2\pi}+g_B\sqrt{g_Dg_F}P^rS_{\bar{r}}|_{\psi=0,2\pi}}{g_F(g_BP^{\varphi}-2g_AP^t)}
    \label{equ:Sbartrphi}
    \end{align}    
\end{subequations}
\end{widetext}
The angle from the spin vector to the $\hat{\bar{r}}$ direction of the observer is then 
\begin{align}
\label{equ:alphadef}
    \alpha_{\hat{\bar{r}}\to \hat{S}}(\psi)\equiv -\arccos\frac{S_{\bar{r}}}{\sqrt{S_{\bar{\mu}}S^{\bar{\mu}}}}
\end{align}
Finally, substituting Eq. \eqref{equ:SbrieSr0} into Eq. \eqref{equ:alphadef}, the precession from one apoapsis at $\psi=0$ to the next at $\psi=2\pi$ is found to be
\begin{align}
\label{eq:alphaaps}
\Delta\alpha_{\hat{\bar{r}}\to \hat{S}}=& \alpha_{\hat{\bar{r}}\to \hat{S}}|_{\psi=2\pi}-\alpha_{\hat{\bar{r}}\to \hat{S}}|_{\psi=0}\nonumber\\
=&-\frac{2\pi\omega_2}{p}+\frac{2\pi\omega_3}{p^{3/2}}-\frac{2\pi\omega_4}{p^{2}}+\mathcal{O}(p^{-5/2})
\end{align}
where $\omega_i$ were given in Eq. \eqref{eq:omegaires}. We also did a parallel computation for the the precession from one periapsis at $\psi=-\pi$ to the next periapsis at $\psi=\pi$. The result can be proven to be exact the same as the above. Henceforth we will not distinguish these two cases. 
Several comments about this result are in order. As seen from $\omega_2$ in Eq. \eqref{eq:omega2res} and the metric expansion in Eq. \eqref{eq:metricexp}, the first two coefficients $a_1,\,a_2$ of $g_A(r)$, representing the ADM mass and total charge of the spacetime, appear at the same order. If the function $g_C(r)$ has a nontrivial expansion coefficient $c_1$, such as for the case of Gibbons–Maeda–Garfinkle–Horowitz–Strominger spacetime \cite{Garfinkle:1990qj,Bhadra:2003zs}, charged dilaton spacetime \cite{Garfinkle:1990qj} and Janis-Newman-Winicour spacetime \cite{Janis:1968zz}, it will also contribute to the leading order of the precession angle. This means that this spin precession angle is very sensitive to the electric or other effective charges of the spacetime. 
The second-order term of the precession angle, at order $p^{-3/2}$, has a coefficient $b_1$, which usually corresponds to the spacetime spin. This suggests that the LT effect on this precession angle appears from the subleading order too, similar to the precession angle Eq. \eqref{eq:omegaslrc}. 

To verify the correctness of this result, next we will study the precession of the spinning object traveling along PN orbit in a KN spacetime, comparing its precession angle  \eqref{eq:alphaaps} obtained perturbatively, with the same angle that is calculated numerical from the original MPD equation.

\subsection{Spin precession along PN orbits in KN spacetime}

To obtain the precession in the KN spacetime, we first asymptotically expand its metric functions in Eq. \eqref{eq:knm} to get
\begin{subequations}
    \label{equ:KN}
    \begin{align}
        g_A(r)=&1-\frac{2M}{r}+\frac{Q^2}{r^2}+\mathcal{O}\left(r^{-3}\right),\\
        g_B(r)=&-\frac{4a M}{r}+\frac{2a Q^2}{r^2}+\mathcal{O}\left(r^{-3}\right),\\
        g_C(r)=&r^2+a ^2+\frac{2a ^2M}{r}-\frac{a ^2Q^2}{r^2}+\mathcal{O}\left(r^{-3}\right),\\
        g_D(r)=&1+\frac{2M}{r}+\frac{4M^2-a ^2-Q^2}{r^2}+\mathcal{O}\left(r^{-3}\right),\\
        g_F(r)=&r^2+\mathcal{O}\left(r^{-3}\right).
    \end{align}
\end{subequations}
Reading off and substituting the expansion coefficients into Eq. \eqref{eq:alphaaps}, the precession of a spin between two neighboring apoapsides in the equatorial plane in the KN spacetime is found to be 
\begin{align}
\Delta\alpha_{\hat{\bar{r}}\to \hat{S},\mathrm{KN}}=&-\frac{(3M^2-Q^2)\pi}{Mp}+\frac{6a \sqrt{M}\pi}{p^{3/2}}-\frac{\pi}{4M^2p^2}\nonumber\\
&\times\Big[12a^2M^2+2(e^2-17)M^2Q^2-Q^4\nonumber\\
&+(63-6e^2)M^4\Big]+\mathcal{O}\left(p^{-5/2}\right).
    \label{equ:c4deltaSKN}
\end{align}
To verify the reliability of Eq. \eqref{equ:c4deltaSKN}, we have compared its value with that from numerical integration of the spin evolutions Eq. \eqref{equ:dhatS2} using a fourth-order Runge-Kutta method. Excellent agreement is found even for $p$ that are not very large (tens of $M$). 

There are a few observations we can make about this precession. The first and most important is that unlike the precession angle in Eq. \eqref{eq:spindirinkn}, here the spacetime charge $Q$ also appears in the leading $p^{-1}$ order. Since when $|Q|>\sqrt{3}M$ this term will change sign, i.e., the spin precession becoming delayed behind the $\hat{\bar{r}}$ direction after returning to the apoapsis. This can also be used as a way to distinguish the NS KN spacetime from the BH one if sign of the precession can be measured. The second is that the spacetime spin effect appears from the second order, which is similar to Eq. \eqref{eq:spindirinkn}. However, the signs of the corresponding terms are different, implying that the corotating object's spin will precess more delayed comparing to the counter-rotating one. The third is that the eccentricity $e$ only shows up from the $p^{-2}$ order, making its effect less apparent in most observation related applications. 
The final point prompts us to investigate the relationship between this precession $\Delta\alpha_{\hat{\bar{r}}\to \hat{S},\mathrm{KN}}$ in Eq. \eqref{equ:c4deltaSKN} and the precession $\alpha_{\hat{x}^+ \to\hat{S},\mathrm{KN}}(\varphi)$ in Eq. \eqref{eq:spindirinkn} as well as the apsis precession angle $\delta_{\mathrm{KN}}$ in Ref. \cite{He:2023joa} (see Eq. (77) therein) in the circular orbit limit.   
From the precession illustrated in Fig. \ref{fig:PN} (b), it is clear that they
should satisfy the relation 
\begin{align}
    &\alpha_{\hat{x}^+ \to\hat{S},\mathrm{KN}}(\varphi=2\pi+\delta_{\mathrm{KN}})-\alpha_{\hat{x}^+ \to\hat{S},\mathrm{KN}}(\varphi=0)\nonumber\\
    &-\Delta\alpha_{\hat{\bar{r}}\to \hat{S},\mathrm{KN}}=\delta_{\mathrm{KN}}.\label{eq:angrel}
\end{align}
We have checked this relation using the above mentioned three angles and found that it holds perfectly, to the highest orders of $p$ in these equations.

\begin{figure}[htp!]
\centering
\includegraphics[width=0.48\textwidth]{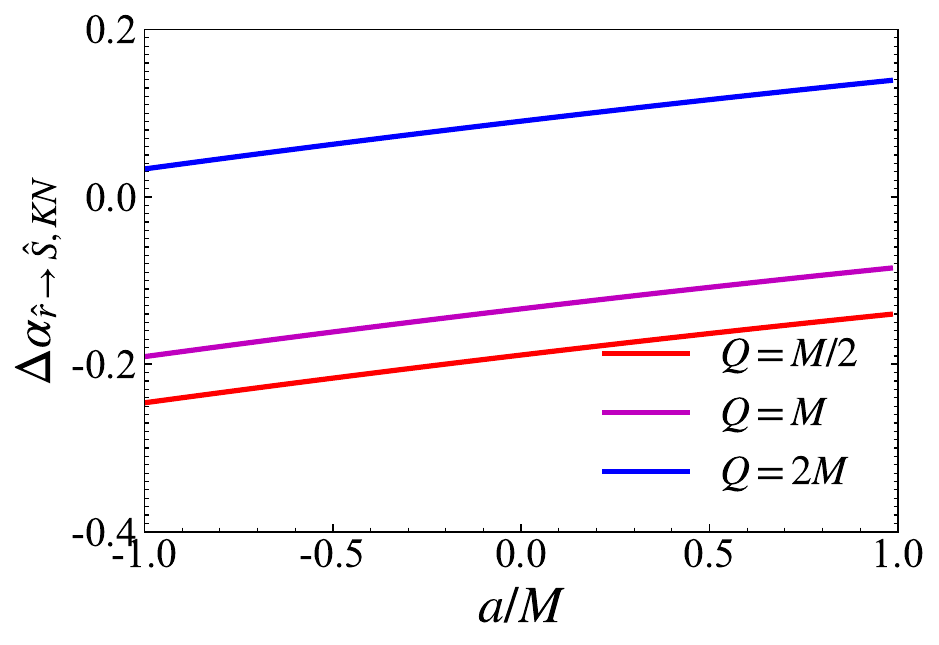}\\
(a)
\includegraphics[width=0.48\textwidth]
{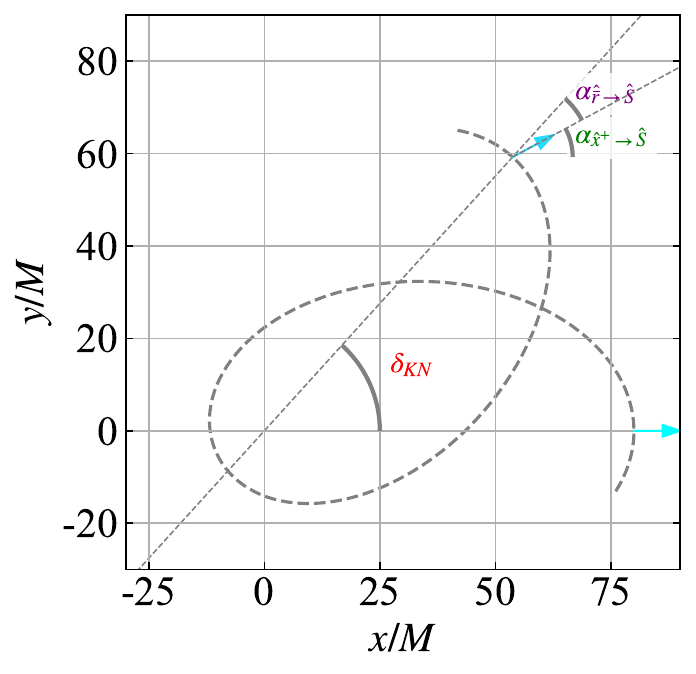}\\
(b)
\caption{Spin precession angle between two neighboring apoapsides, Eq. \eqref{equ:c4deltaSKN}: (a) as a function of $a$ for several $Q$.  $p=100M,\,e=1/2$ in this plot; (b) Precession at the two apoepsides along the trajectory with $p=20M,\, e=1/2,\, Q=M/2,\,a=4M/5$.
  \label{fig:PN}}
\end{figure}

In Fig. \ref{fig:PN} (a), we plot the precession angle \eqref{equ:c4deltaSKN} as a function of $a$ for several $Q$. 
Since we chose a relatively small $p$, the variation as $a$ changes from $-M$ to $M$ is still appreciable although the spin contribution is only at subleading order. The main purpose of this plot is actually to show that this precession angle changes smoothly between the BH and NS parameter choices. Moreover, as seen from the first term of Eq. \eqref{equ:c4deltaSKN}, when $Q>\sqrt{3}M$ in the NS case, the precession can be positive, meaning that the spin precesses ahead of the $\hat{r}$ direction. This will never be achievable if the KN spacetime is a BH one. In plot (b), we plot the precession along a non-circular orbit. In order for the precession angles to be visible, we chose a smaller $p$ and a large $e$. It is easy to understand that the relation \eqref{eq:angrel} should hold, as claimed before.

Before we move to next section to compare our results of precession angles to those astronomical objects, we need to convert these precession angles to the spin precession rate (SPR) because in astronomy, it is the latter that are more often observed. To this end, we will have to divide these precession angles by the corresponding orbital (quasi)-period $T_\mathrm{KN}$, i.e. the time to move around one full $2\pi$ in $\varphi$ coordinate. Since in the spin-orbital plane parallel case, the orbital motion is not affected by the particle spin, this period can be computed using Eq. \eqref{equ:Ppar} and the result to the leading orders in the PN approximation takes the form
\begin{align}
&T_\mathrm{KN}=\int_{r_-}^{r_+}\frac{P^t}{P^r}\dd r\nonumber\\
=&2 \int_{r_-}^{r_+} \frac{E (r^2 + a^2) - a L}{\left[ L \Delta + a E (2Mr - Q^2) \right] \sqrt{E^2 - V(r)}} \frac{r^2}{\sqrt{\Delta}} \, \dd r\nonumber\\
=&\frac{2\pi p^{3/2}}{\sqrt{(1-e^2)M}}\left[\frac{1}{1-e^2}+\frac{3M}{p}-\frac{3a\sqrt{M}}{p^{-3/2}}+\mathcal{O}(p^{-2})\right].
\end{align}
Then combining with Eq. \eqref{eq:angrel} and Eq. (77) of Ref. \cite{He:2023joa}, this yields to the leading three orders the SPR
\begin{align}
&\Omega_\mathrm{KN}\equiv \frac{\delta_\mathrm{KN}+\Delta\alpha_{\hat{\bar{r}}\to \hat{S},\mathrm{KN}}}{T_\mathrm{KN}}\nonumber\\
=&\frac{\sqrt{(1-e^2)^3M}}{p^{5/2}}\Bigg[\frac{3M}{2}-\frac{a\sqrt{M}}{p^{1/2}}\nonumber\\
&+\frac{\left(9+48e^2\right)M^2-(14+4e^2)Q^2}{8p}+\mathcal{O}(p^{-3/2})\Bigg].
\label{eq:sprkn}
\end{align}
For binary system with comparable masses $m_1$ and $m_2$, such as the pulsar binaries listed in Tab. \ref{tab:c4earthperS}, the period should be corrected such that the SPR to the leading order becomes\cite{Barker:1975ae}
\begin{align}
\Omega_\mathrm{KN1}=\frac{3(3m_1+4m_2)\sqrt{(1-e^2)^3m_1^3}}{2(m_1+m_2)p^{5/2}}. \label{eq:omegakn1}
\end{align}

\section{Astrophysical applications \label{sec:app}}

The results derived in Secs. \ref{sec:ssa} and \ref{sec:spinparallel} can be applied to many celestial objects orbiting around a gravitational center, given that most of them carry spin angular momenta and thus experience the precession caused by the gravitational field. For natural systems, the obvious examples include the planet-moon systems, the Sun-planet system, the stars around galaxy centers or other binary systems. Some artificial objects include satellites orbiting the Sun, Earth or other planets, moons, asteroids and comets.  

In general, according to Eq. \eqref{eq:sprkn}, in order to make the precession angle, we seek systems with minimal $p$ and maximal $(M,\, Q,\,a)$. Consequently, most natural satellites do not exhibit larger spin precession angles than artificial orbiters at low altitude around the same central object, as the latter (if any) is typically much closer to central object. However, for many distant systems, artificial satellites are not available and those natural ones are the only possible candidates for verifying the precession angles or SPRs studied in this work. For systems with different central objects (except the compact ones such as BHs or neutron stars), we can estimate the leading-order contribution to the SPR, i.e. $\sim M^{3/2}/p^{5/2}$ using $M\propto \rho R^3$ and $p>R$ with $R$ being the central object's radius. For satellites with low altitude ($p\gtrsim R$), then apparently the larger the $R$, the larger the SPR. 
In general, according to Eq. \eqref{eq:sprkn}, in order to make the precession angle, we seek systems with minimal $p$ and maximal $(M,\, Q,\,a)$. Consequently, most natural satellites do not exhibit larger spin precession angles than artificial orbiters at low altitude around the same central object, as the latter (if any) is typically much closer to central object. However, for many distant systems, artificial satellites are not available and those natural ones are the only possible candidates to verify the precession angles or SPRs studied in this work. Among systems with different central objects (except the compact ones such as BHs or neutron stars), we can estimate the leading order contribution to the SPR, i.e. $\sim M^{3/2}/p^{5/2}$ using $M\propto \rho R^3$ and $p>R$ with $R$ being the central object radius. For satellites with low altitude ($p\gtrsim R$), then apparently the larger the $R$, the larger the SPR. 

In Tab. \ref{tab:c4earthperS}, we list three categories of systems, including the names of the system, corresponding values of $(p,\,e,\,M,\,a)$ and the SPR \eqref{eq:sprkn} to each order, from the left to right columns of the table. The first kind are the natural systems, which for each central object we only present its spinning satellite with (almost) the smallest $p$. This includes three subclasses, the massive BH-star system, the Sun-planet or planet-moon systems, and the binary pulsar systems.  The second category are systems with artificial satellites, again only those with (almost) the smallest $p$ are listed. Some systems in the first two categories with smaller $p$ were not listed because they are perturbed too strongly by the third object (e.g. the Neptune-Naiad system perturbed by Thalassa). In addition to the satellite with smallest $p$, other satellite for the same central object may also be listed, primarily due to their availability of observational data. The third category is for imaginary/test satellites with very low altitude, assumed $1.05R$ (except for Sun which $10R$ is used), around the central bodies. 

\begin{table*}[htp!]
\centering
\begin{tabular}{l|lllll|llll} 
\hline
\hline
Natural systems & $p$ (km)~~~ & $e$~~~~~~~~~~ & $M$ (kg)~~~ & $\hat{a}$ (kg$\cdot$m$^2$/s) & $i$ ($^\circ$) & $p^{-5/2}(^{\prime\prime}$/cy) & $p^{-3}(^{\prime\prime}$/cy) & $p^{-7/2}(^{\prime\prime}$/cy) & Ref. \\ 
\hline
Sgr A*-S2       & $3.343$E10& $8.85$E-1 & $8.475$E36  & $2.390 $E45 & 134.6& 2.277E3      & $+2.19$E-9  &$1.329$E1 &\cite{GRAVITY:2020gka,Moscibrodzka:2009gw}\\
Sun-Mercury     & $5.548$E7 & $2.06$E-1 & $1.989 $E30 & $1.90 $E41  & 3.38 & 2.14E1      & $-5.02$E-4  &$4.19$E-6&\cite{Folkner:2014WM}\\
Earth-Moon      & $3.832$E5 & $5.49$E-2 & $5.972 $E24 & $5.85$E33  & 23.5 & $2.99$E-2 & $-4.58$E-5  &$2.10$E-12&\cite{Moon:2006MA}\\
Mars-Phobos     & $9.374$E3 & $1.51$E-2 & $6.417 $E23 & $1.9 $E32   & 1.093& $1.13$E1      & $-1.11$E-1  &$3.44$E-9&\cite{Phobos:2022KK} \\
Jupiter-Metis   & $1.280$E5 & $2$E-4    & $1.898 $E27 & $4.31$E38 & 0.06 & $2.636$E3      & $-9.91$E1     &$1.74$E-4&\cite{Metis:2004BS}\\
Saturn-Pan      & $1.336$E5 & $1.44$E-5 & $5.683 $E26 & $7.10 $E37  & 1E-4 & $3.88$E2       & $-1.44$E1     &$7.34$E-6 &\cite{Pan:2007RA}\\
Uranus-Cordelia & $4.977$E4 & $2.6$E-4  & $8.681 $E25& $1.261$E36 & 0.085& $2.73$E2       & $-4.93$E0     &$2.12$E-6 &\cite{Cordelia:1998RA} \\
Neptune-Naiad   & $4.822$E4 & $4.7$E-4  & $1.024 $E26 & $1.621$E36 & 0.475& $3.79$E2       & $-6.97$E0     &$3.58$E-6 &\cite{Despina:2004RA} \\
&&&$M$ ($M_\odot$)&$m_1$ ($M_\odot$)&$m_2$ ($M_\odot$)&$p^{-5/2}(^\circ$/yr)& Obs.$(^\circ$/yr) & Diff. \\
PSR B1913+16   & $1.207$E6 & $6.17$E-1  &2.828 & 1.441  & 1.387 & 1.25&1.21&$3.3$\%& \cite{Weisberg:2004hi,Kramer:1998id} \\
PSR J0737-3039   & $8.719$E5 & $8.78$E-2  &2.587&1.338&1.249& 5.06& 4.77&$6.1$\%& \cite{Kramer:2006M,Breton:2008xy}\\
PSR J1906+0746   & $1.201$E6 & $8.5$E-2  &2.623 &1.291&1.332& 2.18& 2.1&   $3.8$\%&\cite{vanLeeuwen:2014sca,Desvignes:2012ak} \\
PSR B1534+12   & $2.112$E6 & $2.74$E-1  &2.679&1.333&1.346& $5.1$E-1& $5.9$E-1&   $13.6$\%&\cite{Fonseca:2014E} \\
PSR J1141-6545  & $1.307$E6 & $1.72$E-1  &2.29&1.27&1.02& 1.69&1.36&$24.2$\%&\cite{Bhat:2008ck,Hotan:2004ub} \\
PSR J1756−2251  & $1.813$E6 & $1.81$E-1  &2571&1341&1230& $7.8$E-1&$7.26$E-1&7.5\%&\cite{Ferdman:2014RD} \\
\hline
Artificial satellites\\
\hline
Sun-Helios1     & $5.795$E7 & $2.58$E-1 & $1.989 $E30 & $1.90$E41  & 0.02  & $1.85$E1      & $-4.25$E-4  &$3.82$E-6 &\cite{HELIOs:1981hp}\\
Sun-Helios2     & $5.665$E7 & $3.06$E-1 & $1.989 $E30 & $1.90$E41  & 0     & $1.87$E1      & $-4.35$E-4  &$4.38$E-6&\cite{HELIOs:1981hp}\\
Sun-PSP         & $1.143$E7 & $8.96$E-1 & $1.989 $E30 & $1.90$E41  & 3.4   & $1.04$E2     & $-5.37$E-3  &$4.25$E-5&\cite{PSP:2020JK}\\
Mercury-Messenger&$4.592$E3 & $7.40$E-1 & $3.301$E23  & $8.53$E29  & 80    & $7.54$E0      & $-2.25$E-4  &$9.46$E-9&\cite{Smith:2012DE}\\
Venus-Magellan  & $8.825$E3 & $3.92$E-1 & $4.868$E24  & $1.80$E31  & 85.5  & $2.13$E2       & $-7.72$E-4  &$9.53$E-7&\cite{Saunders:1992RS}\\
Earth-GPB& $7.027$E3 &$1.4$E-3      & $5.972$E24  & $5.85$E33  & 90     & $6.59$E2       & 0     &$2.49$E-6&\cite{Everitt:2011CF}\\
Moon-Queqiao 2  & $3.969$E3 & $9.75$E-1    & $7.346$E22  & $6.336$E30 & 117    & $4.11$E-2 & $2.52$E-4  &$2.05$E-11&\cite{Hong:2025X}\\
Mars-MRO        & $3.677$E3 & $8.84$E-3    & $6.417 $E23 & $1.9 $E32  & 92.7  & $1.17$E2       & $+8.68$E-2   &$9.10$E-8&\cite{Zurek:2024R}\\
Jupiter-Juno    & $1.495$E5 & $9.74$E-1 & $1.898 $E27 & $4.31$E38 & 90.0  & $2.08$E1      &0&$7.11$E-6&\cite{link:juno}\\
Saturn-Cassini  & $9.445$E5 & $8.96$E-1 & $5.683 $E26 & $7.10$E37  & 12.8  & $2.56$E-1      & $-3.47$E-3    &$3.61$E-9&\cite{link:Cassini}\\
\hline
Test satellites\\
\hline
Sun             & $6.957$E6 &    0      & $1.989 $E30 & $1.90 $E41 & 0     & 4.106E3      & $-2.72$E-1  &$5.22$E-3&\cite{Nyambuya:2015GG}\\
Mercury         & $4.592$E3 &    0      & $3.301$E23  & $8.53$E29  & 0     & 2.48E1       & $-4.25$E-3  &$7.93$E-9&\cite{link:Mercurya}\\
Venus           & $6.354$E3 &    0      & $4.868$E24  & $1.80$E31  & 0     & 6.24E2       & $-3.38$E-2  &$2.12$E-6&\cite{Margot:2021JL}\\
Earth           & $6.69$E3  &    0      & $5.972$E24  & $5.85$E33  & 0     & 7.45E2       & $-9.43$E0       &$2.96$E-6&\cite{link:Mercurya}\\
Mars           & $3.679$E3 &    0      & $6.417 $E23 & $1.9 $E32  & 0     & 1.17E2       & $-1.84$E0       &$9.08$E-8&\cite{Rivoldini:2011A}\\
Jupiter         & $7.34$E4  &    0      & $1.898 $E27 & $4.31$E38 & 0     & $1.06$E4    & $-5.26$E2        &$1.22$E-3&\cite{Iorio:2008mx}\\
Saturn          & $6.11$E4  &    0      & $5.683 $E26 & $7.10 $E37 & 0     & 2.734E3      & $-1.50$E2       &$1.13$E-4&\cite{Helled:2011R}\\
Uranus          & $2.54$E4  &    0      & $8.681 $E25 & $1.261$E36 & 0     & 1.470E3      & $-3.71$E1       &$2.23$E-5&\cite{Nettelmann:2012su}\\
Neptune         & $2.462$E4 &    0      & $1.024 $E26 & $1.621$E36 & 0     & 2.036E3      & $-5.24$E1       &$3.77$E-5&\cite{Nettelmann:2012su}\\
\hline
\hline
\end{tabular}
\caption{Spinning systems and their SPRs using Eq. \eqref{eq:sprkn} or \eqref{eq:omegakn1} with $Q=0$. From left to right: name of the system, the values of $(p,\,e)$ of the satellite, the values of $(M,\,a)$ and inclination angle $i$ of the central object, and the corresponding contribution to the SPR from orders $p^{-5/2},\,p^{-3},\,p^{-7/2}$. The columns and units for the pulsar binary systems are slightly different but self-evident. For Cassini, the orbital data is accurate only for some of its many orbits. Abbreviations: PSP, Parker Solar Probe; GPB, Gravity Probe B; MRO, Mars Reconnaissance Orbiter. References for GPB and the pulsar binaries are for their SPR and/or orbital observational data. Those for the Sgr A*, the Sun and planets are for their spin angular momentum $\hat{a}$. References for launched satellites are for their orbital parameters $p,\,e,\,i$ etc. 
\label{tab:c4earthperS} }
\end{table*}

Among the natural systems in this table, the geodetic spin precession of the pulsars in the binary system are already confirmed and agrees well with the leading order contributions from our computation using Eq. \eqref{eq:omegakn1}. The difference at percentage level can be attributed to the large gravitational torque on the non-spherical shapes of the objects. This torque effect is essentially Newtonian in nature. The sizes of these SPRs are also the largest due to their close distance and high densities. The precession of S2's spin around the Sgr A* is similar in size to the largest SPR in solar system, i.e., the Jupiter-Metis system, which are both at about the $2000^{\prime\prime}$/cy level. Note that the Moon's spin precession around Earth is strongly affected by Sun and therefore the value will receive large correction after solar influence is considered. We did not list the Earth spin precession here although its value is large because primary value (more than 5040.6445$^{\prime\prime}$/cy) of this precession arises from the gravitational effects of the Moon and Sun on Earth's non-ideal spherical shape, while the Schwarzschild-like general relativistic effect is generally considered a second-order contribution ($\sim1.92^{\prime\prime}$/cy, three orders smaller) \cite{souchay2007overview} and the LT effect is even smaller by five orders. A similar situation happens for the Mars spin precession \cite{kahan2021mars,pashkevich2022geodetic}. To our knowledge, none of the spin precessions for these planets or moons in the solar system have been observationally confirmed, primarily due to their small magnitude and observational challenges (such as those of the remote planets/moons). 

In the second category of systems for the artificial satellites that have been launched, the Gravity Probe B was tasked to measure the geodetic spin precession of tops around the Earth. Its measured value of $660.6^{\prime\prime}$/yr matches extremely well with the leading order results from our computation. This order is indeed a Schwarzschild spacetime contribution, without any spin effect of the Earth taken into account. On the other hand, since the Gravity Probe B satellite had a polar orbit, the LT effect on its spin precession is absent at the subleading order (order $p^{-3/2}$) anyways. All other satellites were not specifically designed to measure spin precession. Nevertheless, we still see that the spin precessions of the PSP satellite around the Sun, the Magellan around the Venus and MRO around Mars are all of the same order. 

The last kind shows basically the maximal SPR of a low altitude satellite $1.05R$ from the corresponding planet center or $10R$ from the Sun center. It is seen that the SPR around Jupiter is larger than $\sim 10^{4\prime\prime}$/cy, about 14 times larger than around the Earth, which, in turn, is larger than those around nearby planets, Venus and Mars. Together with the largeness of the SPR of the Jupiter-Metis system, this suggest that close observations of the precession of Jupiter's satellites, either natural or artificial, might fruit earlier in confirming the GR effects in spin precession. Moreover, we also see that for Jupiter, even the SPR at the next order, which is proportional to Jupiter's spin and measures the LT effect, can reach $\sim 530^{\prime\prime}$/cy. This value is well within the detection capability of us using current technology, given that value about 30 times smaller in Gravity Probe B experiment was already discriminated more than two decades ago. 

\section{Conclusions and discussions \label{sec:disc}}

This paper investigates the spin precession of test particles in the equatorial plane of general SAS spacetimes using the MPD equations. We derived the spin precession for two cases: first, when the spin angular momentum per unit mass is much smaller than its own mass $(J/m\ll m)$, and second, when the spin vector is parallel to the orbital plane. Under the small-spin approximation, we showed that the component perpendicular to the equatorial plane, $S^\theta$, remains constant along motion. The other two components $(S^r,\,S^\varphi)$ were solved for circular orbits, and the spin precession angle after each circle in genearl SAS spacetime is found in Eqs. \eqref{equ:omegasdef}, \eqref{eq:omegaslrc} and \eqref{eq:alphacirc}. Their specific forms in the KN spacetime, i.e. Eqs. \eqref{eq:omegasasy} and \eqref{eq:spindirinkn}, were studied in detail. For the spin-orbital plane parallel case, we found that under this assumption the particle's orbital motion completely decouples from its spin evolution. Although the equations of motion for the spin components $(S^r,\,S^\varphi)$ in this case cannot be solved analytically in general, we were still able to show that a special variable transformation allows these equations to decouple from each other completely. Using the PN approximation to the resultant equations, we derived the spin precession angle between two consecutive apoapsides as a power series of $p^{-1/2}$ to an arbitrarily high order in Eq. \eqref{eq:alphaaps}. In the KN spacetime, as shown in Eq. \eqref{equ:c4deltaSKN} the charge $Q$ appears alongside the mass $M$ in the leading order, while the spacetime spin $a$ appears only in the next order. The orders in which these spacetime parameters appear are similar to those in the periapsis shift angles \cite{He:2023joa,Xu:2024msu}. 

We then applied the spin precession rates to astronomical systems with spinning secondaries, including the Sgr A*-S2 system, some binary pulsar systems with known spin precession, the Sun-planet and planet-moon systems, as well as Sun/planet/moon-artificial satellite systems. The SPRs computed using our formula agree well with systems having observational data, including binary pulsar systems and the GPB satellite. The SPRs for other systems can be considered prediction from our results. we found that the SPR around Jupiter is the largest among the near-by systems surveyed, and even its second order, which can test the LT effect, should be measurable by technologies used in GPB satellite. 

Several comments about the work and its extension are warranted here. First, although we applied our precession angle formulas only to the KN spacetime, they are derived for general SAS and thus can be easily applied to other SAS spacetimes to reveal the effects of spacetime parameters of interest on spin precession. This is particularly important spacetimes with both BH and NS scenarios since our result provides an observational tool to distinguish them. Secondly, this work examines the motion of test particles solely in the equatorial plane. A generalization to non-equatorial orbits would be valuable as we expect that the particle spin may exhibit non-trivial coupling with the central body's spin when the orbit is not planar. Finally, the current analysis considers only the motion of spinning particles in a gravitational field. However, if both the central body and the test particle carry electric charges, electromagnetic effects would arise. 

\section*{Acknowledgements}

S. Xu is supported by the Undergraduate Training Program for Innovation of Wuhan University (S202410486082). This work is partially supported by a research development fund from Wuhan University. 

\appendix
\section{Solution to Eq. \eqref{equ:dSpar} \label{app:eqsol}}

In this appendix, we solve \eqref{equ:dSpar} using the PN method. We first show that this coupled system of $(S_r,\, S_\varphi)$ can be separated via a linear transformation, allowing the resultant equations be solved perturbatively. Then we transform the solution back to that to $(S_r,\, S_\varphi)$.

The linear transformation we propose from $(S_r,\, S_\varphi)$ to $(\hat{S}_r,\,\hat{S}_\varphi)$ is defined through the following relation  \cite{Pang:2024tco}
\begin{subequations}
    \label{equ:tranhatS}
    \begin{align}
        S_r=&\frac{g_D\hat{S_r}}{f(r)},\\
        S_{\varphi}=&\frac{g_BP^{r}g_D}{2Ef(r)}\hat{S_r}+\frac{Lg_B+2Eg_C}{2Eh(r)}\hat{S_{\varphi}}
    \end{align}
\end{subequations}
where $f(r)$ and $h(r)$ satisfy the equations
\begin{subequations}
\label{equ:dfh}
    \begin{align}
        \frac{h^{\prime}(r)}{h(r)}=&\frac{g_F^{\prime}}{g_F}-\frac{g_B\left(Eg_B^{\prime}-2Lg_A^{\prime}\right)+2g_A\left(2Eg_C^{\prime}+Lg_B^{\prime}\right)}{4g},\\
        \frac{f^{\prime}(r)}{f(r)}=&\frac{1}{2}\left(\frac{g_D^{\prime}}{g_D}+\frac{g_F^{\prime}}{g_F}+\frac{2p^tg_A^{\prime}-p^{\varphi}g_B^{\prime}}{2E}\right).
    \end{align}
\end{subequations}
Note that each of Eqs. \eqref{equ:dfh} is a separable first-order ODE and therefore $f(r)$ and $h(r)$ can be readily obtained from a differential equation perspective. This implies that once $(\hat{S}_r,\,\hat{S}_\varphi)$ in Eq. \eqref{equ:tranhatS} are known, they allow us to determine $(S_r,\,S_\varphi) $ immediately. 

Substituting Eq. \eqref{equ:tranhatS} into \eqref{equ:dSpar}, we obtain an antidiagonal ODE system of the form
    \begin{align}
    \label{equ:dhatS}
\begin{pmatrix}
    \hat{S}_r^\prime \\
    \hat{S}_\varphi^\prime 
\end{pmatrix}        
=\begin{pmatrix}
        0&\Omega_1(r)\\
        \Omega_2(r)&0\\
    \end{pmatrix}
    \begin{pmatrix}
        \hat{S}_r\\
    \hat{S}_\varphi 
    \end{pmatrix}
    \end{align}
where the elements of the coefficient matrix are
\begin{subequations}
\label{equ:Omega12}
\begin{align}
    \Omega_1(r)=&\frac{f(r)}{h(r)Eg_D\left(g_B^2+4g_Ag_C\right)P^r}\Big[E^2\left(g_Cg_B^{\prime}-g_Bg_C^{\prime}\right)\nonumber\\
        &+L^2\left(g_Ag_B^{\prime}-g_Bg_A^{\prime}\right)+2EL\left(g_Ag_C^{\prime}-g_Cg_A^{\prime}\right)\Big],\\
         \Omega_2(r)=&\frac{h(r)}{f(r)E\left(g_B^2+4g_Ag_C\right)^2P^r}\bigg\{\Big[m^2(g_B^2+4g_Ag_C)\nonumber\\
         &+4L^2g_A-4ELg_B-4E^2g_C\Big]\left(g_Ag_B^{\prime}-g_A^{\prime}g_B\right)\nonumber\\
         &+\Big[\left(2ELg_A^{\prime}-E^2g_B^{\prime}\right)\left(g_B^2+4g_Ag_C\right)\Big]^{\prime}\bigg\}.
    \end{align}
\end{subequations}
Differentiating Eq. \eqref{equ:dhatS} and reusing it, we finally decouple
$\hat{S}_r$ and $\hat{S}_\varphi$
\begin{subequations}
        \label{equ:dhatS2}
    \begin{align}
        \frac{\dd^2\hat{S}_r}{\dd r^2}=&\frac{\Omega_1^{\prime}}{\Omega_1}\frac{\dd\hat{S}_r}{\dd r}+\Omega_1\Omega_2\hat{S}_r, \label{equ:dhatS2a}\\
        \frac{\dd^2\hat{S}_{\varphi}}{\dd r^2}=&\frac{\Omega_2^{\prime}}{\Omega_2}\frac{\dd\hat{S}_{\varphi}}{\dd r}+\Omega_1\Omega_2\hat{S}_{\varphi}.\label{equ:dhatS2b}
    \end{align}
\end{subequations}
Note that an inspection can show that these equations depend on $f(r)$ and $h(r)$ solely through the ratios $f^{\prime}(r)/f(r)$ and $h^{\prime}(r)/h(r)$, not $f(r)$ and $h(r)$ themselves. These ratios can be directly quoted from Eq. \eqref{equ:dfh} so we do not have to solve for them when solving $\hat{S}_r$ and $\hat{S}_{\varphi}$. However, when transforming from $(\hat{S}_r,\,\hat{S}_\varphi)$ back to $(S_r,\,S_\varphi) $, we do require the explicit forms of $h(r)$ and $f(r)$. One more point is that Eqs. \eqref{equ:dhatS2} are homogeneous so that their solutions allow a constant factor difference unless proper initial conditions are applied.

Now Eq. \eqref{equ:dhatS2}, although decoupled, is still a second order ODE system with variable coefficients, and thus no general solution in closed-form can be obtained immediately. To solve them, as noted in Sec. \ref{subsec:smallpgp}, we apply the PN approximation, assuming a large orbital radius for the spinning object. That is, the semi-latus rectum $p$ in Eq. \eqref{equ:pnorbit} is large compared to the mass parameter of the spacetime. Next we will first simplify the Eqs. \eqref{equ:dhatS2} and then solve them. We will illustrate this simplification and solution process for Eq. \eqref{equ:dhatS2a} alone. The process for Eq. \eqref{equ:dhatS2b} is similar.

We first change the variables in Eq. \eqref{equ:dhatS2a} from $r$ to $\psi$ using Eq. \eqref{equ:pnorbit} 
\begin{align}
            &\frac{(1+e\cos\psi)^4}{e^2p^2\sin^2\psi}\frac{\dd^2\hat{S}_{r}}{\dd\psi^2}=\Omega_1\Omega_2+\frac{(1+e\cos\psi)^2}{ep\sin\psi}\left\{\frac{\Omega^{\prime}_1}{\Omega_1}\right.\nonumber\\
            &\left.+\frac{1+e\cos\psi}{ep\sin^2\psi}(\cos\psi+e\cos^2\psi+2e\sin^2\psi)\right\}\frac{\dd\hat{S}_{r}}{\dd\psi}.
        \label{equ:dhatSrpsi}
\end{align}
In this equation, the terms containing $\Omega_1$ and $\Omega_2$, as shown in Eq. \eqref{equ:dhatS2}, are still functions of the metric functions and motion constants $(E,\,L)$ and thus require proper treatment under the PN approximation. To this end, we assume that the metric functions admit asymptotic expansions as given in \eqref{eq:metricexp}.
These expansions can then be substituted into Eq. \eqref{equ:dhatSrpsi} and further expanded for large $p$. Before that however, for the $(E,\,L)$ in $\Omega_1$ and $\Omega_2$, we can link them to the kinetic variables $(p,\,e)$ directly from the orbital Eq. \eqref{equ:Pparr}, \eqref{equ:upar} and the conditions $u^{r}|_{r=r_\pm}=0$ with Eq. \eqref{eq:r1r2def}. 
Assuming that under the PN approximation, $(E,\,L)$ have expansions of the form
    \begin{align}
        E=\sum_{i=0}\frac{E_i}{p^{i/2}},\quad
        L=\sum _{i=-1}\frac{L_i}{p^{i/2}},
    \end{align}
we can solve these expansion coefficients from the above equations and to obtain to the first few orders
\begin{subequations}
\label{equ:ELpexpansion}
    \begin{align}
        E=&1+\frac{a_1(1-e^2)}{4p}+\frac{a_1(3a_1-4c_1)(1-e^2)}{32p^2}+\mathcal{O}(p^{-5/2})\\
        L=&\sqrt{\frac{-a_1p}{2}}-\frac{4a_2-a_1(a_1-c_1)(3+e^2)}{4\sqrt{-2a_1p}}-\frac{b_1(3+e^2)}{4p}\nonumber\\
        &+\mathcal{O}(p^{-3/2}).
    \end{align}
\end{subequations}
Note that as the orbital motion in this case is unaffected by the spin, the relations \eqref{equ:ELpexpansion} are identical to those obtained in Ref. \cite{He:2023joa}.

Finally, substituting the metric Eqs. \eqref{eq:metricexp} and \eqref{equ:ELpexpansion} for $(E,\,L)$ into Eq. \eqref{equ:dhatSrpsi}, and performing the large-$p$ expansion transforms this equation into
    \begin{align}
        &\frac{\dd^2 \hat{S}_{r}}{\dd\psi^2}+\hat{S}_r=\left[\frac{3a_1^2-2a_2-a_1c_1-a_1(a_1-2c_1)e\cos\psi}{2a_1p}\right.\nonumber\\
        &\left.+\frac{b_1(-3e\cos\psi)}{\sqrt{-2a_1p^3}}+\mathcal{O}\left(p^{-2}\right)\right]\hat{S}_r+\left[\frac{(-2a_1+c_1)}{2p} e\sin\psi\right.\nonumber\\
        &\left.-\frac{b_1e\sin\psi}{\sqrt{-8a_1p^3}}+\mathcal{O}\left(p^{-2}\right)\right]\frac{\dd \hat{S}_r}{\dd\psi}.
    \label{dSr2SrP}
    \end{align}
We assume the solution takes the form
\begin{align}
\hat{S}_r(\psi)=c_r\left[\hat{S}_{r0}(\psi,\eta(\psi))+\sum_{i=2}\frac{1}{p^{i/2}}\hat{S}_{ri}(\psi,\eta(\psi))\right]
    \label{equ:c4hatSrp}
\end{align}
where we assume that the series function $\hat{S}_{ri}$ depends on $\psi$ in two ways: explicitly and  through an implicit function $\eta(\psi)$. Note that in the summation above, we omit the $p^{-1/2}$ order as it turns out that this order is absent. Eq. \eqref{equ:c4hatSrp} can be solved by substitution into Eq. \eqref{dSr2SrP} and matching each order with appropriate initial conditions. The detailed solving process for ansatz \eqref{equ:c4hatSrp} is given in Appendix \ref{app:eqsol2} and the final solution to the first three orders is 
\begin{subequations}
\label{equ:Srpexp}
    \begin{align}
        \hat{S}_{r0}=&\sin\left(\psi+\eta+\psi_{r0}\right),\\
        \hat{S}_{r2}=&\frac{(2c_1+ea_1)e  }{4}\sin(\psi+\psi_{r0})\sin\eta +e\big[c_1\sin\left(\eta+\psi_{r0}\right)\nonumber\\
       &+\left(c_1-a_1\right)\sin\left(\psi+\eta+\psi_{r0}\right)\big]\sin^2\frac{\psi}{2}, \\
       \hat{S}_{r3}=&-\frac{eb_1}{\sqrt{-8a_1}}\left\{ \cos(\psi-\psi_{r0})\sin\eta \right. \nonumber\\
       &\left.+\sin\psi \left[\cos\left(\eta+\psi_{r0}\right)-\cos\left(\psi+\eta+\psi_{r0}\right)\right]\right\}.
    \end{align}
\end{subequations}
where
\begin{align}
\label{equ:dualscale}
\eta=\sum_{i=2}\frac{\omega_i}{p^{i/2}}\psi
\end{align}
with its first few coefficients
\begin{subequations}
\label{eq:omegaires2}
    \begin{align}
        \omega_2=&-\frac{3a_1^2-4a_2-2a_1c_1}{4a_1},\\
        \omega_3=&-\frac{3b_1}{\sqrt{-8a_1}},\\
        \omega_4=&-\frac{1}{32a_1^2}\Big[\left(6e^2-63\right)a_1^4+(56c_1-16e^2)a_1^3\nonumber\\
        &+\left(8e^2c_1^2-48c_2-8e^2a_2+136a_2\right)a_1^2\nonumber\\
        &+\left(-96a_3+16a_2c_1-64a_2c_1\right)a_1+16a_2^2\Big].
    \end{align}
\end{subequations}
Note that in the solution \eqref{equ:Srpexp}, the only free constants are the amplitude $c_r $ and the initial phase $\psi_{r0}$ while all other variables are determined by the metric functions and the kinetic variables $(p,\,e)$. 

The same method used from Eq. \eqref{equ:dhatSrpsi} to \eqref{equ:c4hatSrp} can be applied to solve $\hat{S}_{\varphi}$ from Eq. \eqref{equ:dhatS2b} yielding
\begin{align}
    \hat{S}_\varphi(\psi)=c_\varphi \left[ \hat{S}_{\varphi 0}(\psi,\eta(\psi))+\sum_{i=2}\frac{1}{p^{i/2}}\hat{S}_{\varphi i}(\psi,\eta(\psi))\right]
    \label{equ:c4hatSphip}
\end{align}
with the first three orders given by
\begin{subequations}
\label{equ:Sphipexp}
    \begin{align}
        \hat{S}_{\varphi0}=&\sin\left(\psi+\eta+\psi_{\varphi 0}\right),\\
        \hat{S}_{\varphi2}=&-\frac{\left[-2c_1+\left(2+e\right)a_1\right]e}{4} \cos(\psi-\psi_{\varphi0})\sin\eta \nonumber\\
        &+e\big[c_1\sin\left(\psi+\eta+\psi_{\varphi 0}\right)\nonumber\\
       &+\left(c_1-a_1\right)\sin\left(\eta+\psi_{\varphi 0}\right)\big]\sin^2\frac{\psi}{2}, \\
       \hat{S}_{\varphi3}=&-\frac{eb_1}{\sqrt{-8a_1}}\left\{\cos(\psi-\psi_{\varphi0})\sin\eta\right.\nonumber\\
&\left.+\sin\psi\left[\cos\left(\eta+\psi_{\varphi 0}\right)-\cos\left(\psi+\eta+\psi_{\varphi 0}\right)\right]\right\}.
    \end{align}
\end{subequations}
Furthermore, using the large $p$  
expansion of the metric functions and $(E,\,L)$ as in Eqs. \eqref{eq:metricexp} and \eqref{equ:ELpexpansion}, we can readily solve Eq. \eqref{equ:dfh} with the initial condition $f(\psi=0)=h(\psi=0)=1$ to obtain
\begin{subequations}
\label{equ:serhf}
\begin{align}
    h(\psi)=&q^2\left\{1+\frac{1}{2a_1p}\left[4\ln q (a_1^2-a_2-a_1c_1)\right.\right.\nonumber\\
    &\left.-ea_1(a_1-c_1+2d_1+f_1)(1-\cos\psi)\right]\nonumber\\
    &\left.+\frac{4b_1\ln q}{\sqrt{-2a_1}p^{3/2}}+\mathcal{O}(p^{-2})\right\},\\
    f(\psi)=&q\left\{1+\frac{1}{2a_1 p}\left[2\ln q (a_1^2-a_2-a_1c_1)\right.\right.\nonumber\\
    &\left.-ea_1(a_1-c_1+2d_1+f_1)(1-\cos\psi)\right]\nonumber\\
    &\left.+\frac{2b_1\ln q}{\sqrt{-2a_1}p^{3/2}}+\mathcal{O}(p^{-2})\right\},\\
    q=&\frac{1+e}{1+e\cos\psi}.
\end{align}
\end{subequations}

Note that when transforming from Eq. \eqref{equ:dSpar} for $(S_r,\, S_\varphi)$ to Eq. \eqref{equ:dhatS2} for $(\hat{S}_r,\,\hat{S}_\varphi)$, the order of each equation increased by one because each is differentiated once. This introduced two additional integration constants in the solutions \eqref{equ:c4hatSrp} and \eqref{equ:c4hatSphip}. Thus, we establish two additional constraints on the four constants $(c_r,\,c_\varphi,\,\psi_{r0},\,\psi_{\varphi0})$, effectively reducing the number of independent initial condition parameters to two. 
First, we consider the initial values of $(S_r(\psi),\,S_{\varphi}(\psi))$ at $\psi=0$, i.e., at the apoapsis. Denoting these values as $(S_{r0},\,S_{\varphi0})$, and noting from Eqs. \eqref{equ:c4hatSrp}, \eqref{equ:c4hatSphip} that the initial values satisfy
\begin{align}
&\hat{S}_r(\psi=0)=c_r\sin\psi_{r0},\,\hat{S}_{\varphi}(\psi=0)=c_\varphi\sin\psi_{\varphi0},
\end{align}
we evaluate Eq. \eqref{equ:dSpar} at the apoapsis to obtain
\begin{subequations}
\label{equ:transS0}
    \begin{align} S_{r0}=&\frac{g_D\hat{S}_r(\psi)}{f(\psi)}\bigg|_{\psi=0}= g_D c_r \sin\psi_{r0},\\ S_{\varphi0}=&\frac{g_BP^{r}g_D}{8Eg_0f(\psi)}\hat{S}_r(\psi)+\frac{Lg_B+2Eg_C}{8Eg_0h(\psi)}\hat{S}_{\varphi}(\psi)\bigg|_{\psi=0}\nonumber\\ =&\frac{g_BP^{r}g_D}{8Eg_0}c_r \sin\psi_{r0}  +\frac{Lg_B+2Eg_C}{8Eg_0}c_{\varphi}\sin\psi_{\varphi0} 
    \end{align}
\end{subequations}
where all metric functions are evaluated at $r=r_+=p/(1+e)$. We establish another set of relations by studying the initial value of the first derivatives of $(\hat{S}_r(\psi),\,\hat{S}_{\varphi}(\psi))$ using the definition 
\begin{subequations}
    \begin{align}
    \frac{\dd\hat{S}_{r}}{\dd \psi}\bigg|_{\psi=0}=&\frac{\dd\hat{S}_{r}}{\dd r}\frac{\dd r}{\dd\psi}\bigg|_{\psi=0}\\
    \frac{\dd \hat{S}_{\varphi}}{\dd \psi}\bigg|_{\psi=0}=&\frac{\dd \hat{S}_{\varphi}}{\dd r}\frac{\dd r}{\dd \psi}\bigg|_{\psi=0}.
    \end{align}
\end{subequations}
Substituting solutions \eqref{equ:c4hatSrp} and \eqref{equ:c4hatSphip} into the left-hand sides and using Eqs. \eqref{equ:dhatS} and \eqref{equ:pnorbit} for the right-hand sides, these equations reduce to
\begin{widetext}
\begin{subequations}
\label{equ:dhatS0}
    \begin{align}
& \left[1+\frac{\omega_2}{p}+\frac{\omega_3}{p^{3/2}}+\mathcal{O}(p^{-2})\right]c_{r}\cos\psi_{r0}=\Omega_1\hat{S}_{\varphi}\frac{\dd r}{\dd\psi}\Big|_{\psi=0}\nonumber\\
=&
\Bigg\{1+\frac{2a_2+a_1\Big[2c_1-(1-e)a_1-(1+e)d_1\Big]}{2a_1p}-\frac{(3-e)b_1}{2\sqrt{-2a_1}}+\mathcal{O}(p^{-2})\Bigg\}\frac{p}{1+e} c_\varphi \sin\psi_{\varphi 0},\\
& \left[1+\frac{\omega_2}{p}+\frac{\omega_3}{p^{3/2}}+\mathcal{O}(p^{-2})\right]c_{\varphi}\cos\psi_{\varphi0}=\Omega_{2}\hat{S}_{r}\frac{\dd r}{\dd\psi}\Big|_{\psi=0}\nonumber\\
=&-\Bigg\{1+\frac{2a_2-a_1\Big[2a_1+(d_1-2ec_1+ed_1)\Big]}{2a_1p}-\frac{(3-e)b_1}{2\sqrt{-2a_1}}+\mathcal{O}(p^{-2})\Bigg\}\frac{1+e}{p}c_r\sin\psi_{r0}
    \end{align}
\end{subequations}
Solving Eqs. \eqref{equ:transS0} and \eqref{equ:dhatS0}, the following combinations of the original constants are expressed in terms of the two new constants $S_{r0}$ and $S_{\varphi0}$
\begin{subequations}
    \begin{align}
        c_r \sin\psi_{r0}=&\bigg[1-\frac{(1+e)d_1}{p}+\frac{(1+e)^2(d_1^2-d_2)}{p^2}+\mathcal{O}(p^{-5/2})\bigg]S_{r0}\\ c_{\varphi}\sin\psi_{\varphi0}=&\bigg[1-\frac{(1+e)c_1}{p}+\frac{(1+e)^2(c_1^2-c_2)}{p^2}+\mathcal{O}(p^{-5/2})\bigg]\frac{(1+e)^2S_{\varphi0}}{p^2}\\ c_r \cos\psi_{r0}=&\bigg[1+\frac{(1+2e)(a_1-2c_1-2d_1)}{p} +\frac{eb_1}{2\sqrt{-2a_1}p^{3/2}}+\mathcal{O}(p^{-2})\bigg]\frac{(1+e)S_{\varphi0}}{p}\\ c_{\varphi}\cos\psi_{\varphi0}=&-\bigg[1+\frac{\Big\{7a_1^2+2a_1\big[(1+e)d_1-(1-2e)c_1\big]-8a_1\Big\}}{4a_1p} -\frac{(6-e)b_1}{2\sqrt{-2a_1}p^{3/2}}+\mathcal{O}(p^{-2})\bigg]\frac{(1+e)S_{r0}}{p}
    \end{align}
\end{subequations}
\end{widetext}

Note that these combinations of $(c_r,\,c_\varphi,\,\psi_{r0},\,\psi_{\varphi0})$ are precisely what is needed in Eqs. \eqref{equ:c4hatSrp} and \eqref{equ:c4hatSphip} to express $\hat{S}_r$ and $\hat{S}_\varphi$ solely in terms of the initial constants $S_{r0}$ and $S_{\varphi0}$. Substituting these, as well as the solutions \eqref{equ:c4hatSrp} and \eqref{equ:c4hatSphip} into Eq. \eqref{equ:tranhatS}, the solutions to $(S_r,\,S_\varphi)$ are obtained as in Eq. \eqref{equ:srsphisol}. 

\section{Method of multiple scales to the solution of Eq. \eqref{dSr2SrP} \label{app:eqsol2}}

When incorporating perturbation effects into periodic functions, one must be cautious, as perturbations can alter the fundamental period of the function. This alteration, even small, could lead to divergent amplitudes or phase shifts if conventional perturbation methods are applied directly. To address this issue, we propose a novel perturbation scheme based on the multiple-scale method. This method introduces two distinct time scales: the intrinsic, fast periodic scale $\xi$, which equals $\psi$ here, and the slowly varying scale $\eta(\psi)$
\begin{align}
\label{equ:dualscale}
\xi=&\psi,\nonumber\\
\eta=&\sum_{i=2}\frac{\omega_i}{p^{i/2}}\psi
\end{align}
where $\omega_i$ represents a series of undetermined coefficients, with the leading-order term assumed to be
\begin{align}
    \omega_2=\frac{3a_1^2-4a_2-2a_1c_1}{-4a_1}.
\end{align}
This selection is made to simplify subsequent notation. In actual computations, choosing any non-zero value for $\omega_2$ will not affect the final results. Under this multiple-scale description in Eqs. \eqref{equ:dualscale}, the system's evolution is now characterized by the differential operator
\begin{align}
    \frac{\dd}{\dd\psi}=\frac{\partial}{\partial \xi}+\sum_{i=2}\frac{\omega_i}{p^{i/2}}\cdot \frac{\partial}{\partial \eta}.
    \label{equ:c4dpsi}
\end{align}
In addition to implementing the aforementioned perturbation treatment to the system's scales, we similarly apply perturbation analysis to the target functions $\hat{S}_r(\psi)$, assuming it takes the form of Eq. \eqref{equ:c4hatSrp}, i.e.,
\begin{align}
\hat{S}_r(\psi)=c_r\left[\hat{S}_{r0}(\xi,\eta)+\sum_{i=2}\frac{1}{p^{i/2}}\hat{S}_{ri}(\xi,\eta)\right].
    \label{equ:c4hatSrprep}
\end{align}

Substituting both Eqs. \eqref{equ:c4dpsi} and \eqref{equ:c4hatSrprep} into Eq. \eqref{dSr2SrP}, we then expand the equation in powers of $p$. The undertermined coefficient method then yields an iterative set of equations
\begin{subequations}
    \label{equ:dhatSseries}
    \begin{align}
        \frac{\partial^2}{\partial \xi^2}\hat{S}_{r0}+\hat{S}_{r0}=&0,\\
        \frac{\partial^2}{\partial \xi^2}\hat{S}_{r2}+\hat{S}_{r2}=&\frac{3a_1-4a_2-2a_1c_1-a_1(a_1-2c_1)e\cos\xi}{2a_1}\hat{S}_{r0}\nonumber\\
        &-\frac{2a_1-c_1}{2}e\sin\xi\frac{\partial}{\partial\xi}\hat{S}_{r0}\nonumber\\
        &+\frac{3a_1-4a_2-2a_1c_1}{2a_1}\frac{\partial^2}{\partial\xi\partial\eta}\hat{S}_{r0},\\
        \frac{\partial^2}{\partial \xi^2}\hat{S}_{r3}+\hat{S}_{r3}=&\frac{b_1(3-e\cos\xi)}{\sqrt{-2a_1}}\hat{S}_{r0}-\frac{eb_1\sin\xi}{\sqrt{-8a_1}}\frac{\partial}{\partial\xi}\hat{S}_{r0}\nonumber\\
        &-2\omega_3\frac{\partial^2}{\partial\xi\partial\eta}\hat{S}_{r0}.\\
        \dots\nonumber
    \end{align}
\end{subequations}
The initial conditions for these equations are specified such that at $\psi=0$
\begin{align}
\hat{S}_{r0}|_{\psi=0}=&\frac{1}{c_r}\hat{S}_{r}\Big|_{\psi=0},\quad \frac{\dd}{\dd\xi}\hat{S}_{r0}\Big|_{\psi=0}=\frac{1}{c_r}\frac{\dd}{\dd\psi}\hat{S}_{r}\Big|_{\psi=0},\\
\hat{S}_{ri}(0,\eta)=&A_i(\eta),\quad \frac{\partial}{\partial \xi}\hat{S}_{ri}(0,\eta)=B_i(\eta),\quad  i=2,3,\cdots. \label{eq:cacbdef}
\end{align}
Here $\hat{S}_{r}\big|_{\psi=0}=c_r\sin \psi_{r0}$ is assumed and $A_i$ and $B_i$ are undetermined functions that depend solely on the slow perturbation scale $\eta$, with the initial condition that they equal $0$ at $\eta=0$.

In brief, we recognize that we have introduced the following unknown quantities: $\omega_i,\,\hat{S}_{ri},\,A_i,\,B_i$. We now present a systematic recursive method to determine these unknowns. First, for the unknowns $A_i,\,B_i,\,\omega_i$, their constraints are determined by the solvability conditions for equations of $\hat{S}_{ri}$. These conditions arise from requiring non-resonant solutions to oscillatory-type differential equations of the form
\begin{align}
    \frac{\partial^2\hat{S}_{ri}}{\partial\xi^2}+\hat{S}_{ri}=\sum_{k=0}\left[f_k(\eta)\cos (k\xi)+g_k(\eta)\sin(k\xi)\right].
\end{align}
If the right-hand side coefficients $f_k,\,g_k\neq0$ for $k=1$, the solution would include secular terms of the form $(\xi\sin\xi,\,\xi\cos\xi)$, which diverge linearly with $\xi$. To eliminate this unphysical behavior, the solvability condition requires that all resonant driving terms (those matching the natural frequency of the homogeneous system) in the differential equation must vanish. Applying these conditions {\it iteratively} to Eq. \eqref{equ:dhatSseries} of $\hat{S}_{ri}$ and using definitions \eqref{eq:cacbdef}, we obtain, for each order $i$, a small system of ODEs of $A_i$ and $B_i$
\begin{subequations}
\label{eq:dadbdeta}
    \begin{align}
        \frac{\dd}{\dd\eta}A_i=&-B_i+\mathcal{A}_i(\eta,\omega_{j+2}),\\
        \frac{\dd}{\dd\eta}B_i=&A_i+\mathcal{B}_i(\eta,\omega_{j+2}),\\
        &i=2,\,3,\,\cdots,\quad j=0,\,1,\,\cdots,\,i;
    \end{align}
\end{subequations}
where $\mathcal{A}_i$ and $\mathcal{B}_i$ are finite sums of $\sin(k\eta)$ and $\cos(k\eta)~(k=1,2,3,\cdots)$ with coefficients being linear combinations of $\omega_j$ upto $\omega_{i+2}$.  Differentiating Eq. \eqref{eq:dadbdeta} again with respect to $\eta$, a new set of equations can be obtained. From the solvability conditions of this new set of equations, one can determine all $\omega_i$ simultaneously, with their solutions presented in Eq. \eqref{eq:omegaires2}, and subsequently the solutions $A_i$ and $B_i$ to Eq. \eqref{eq:dadbdeta}. 
Substituting these $\omega_i$ back into Eq. \eqref{equ:dhatSseries}, we can solve these differential systems for $\hat{S}_{ri}$ with the known initial condition with $(A_i,\,B_i)$ obtained in last step. The solutions for $\hat{S}_{ri}$ are presented in Eq. \eqref{equ:Srpexp}, or equivalently in Eq. \eqref{eq:srpsisol}.

\end{document}